# Layered topological semimetal GaGeTe: new polytype with non-centrosymmetric structure

S. Gallego-Parra,[1] E. Bandiello,[1] A. Liang,[2] E. Lora da Silva,[3] P. Rodríguez-Hernández,[4] A. Muñoz,[4] S. Radescu,[4] A.H. Romero,[5] C. Drasar,[6] D. Errandonea,[2] and F. J. Manjón[1,*]

[1] Instituto de Diseño para la Fabricación y Producción Automatizada, MALTA Consolider Team, Universitat Politècnica de València, 46022 Valencia, Spain

[2] Departamento de Física Aplicada-ICMUV, MALTA Consolider Team, Universitat de València, 46100 Burjassot, Spain

[3] IFIMUP, Departamento de Física e Astronomia, Faculdade de Ciências, Universidade do Porto, 4169-007 Porto, Portugal

[4] Departamento de Física, Instituto de Materiales y Nanotecnología, MALTA Consolider Team, Universidad de La Laguna, La Laguna, 38205 Tenerife, Spain

[5] Department of Physics and Astronomy, West Virginia University, Morgantown, West Virginia 26506-6315, USA

[6] Faculty of Chemical Technology, University of Pardubice, Pardubice 532 10, Czech Republic

* corresponding author: fjmanjon@fis.upv.es

## Abstract

GaGeTe is a layered material composed of germanene and GaTe sublayers that has been recently predicted to be a basic $Z_2$ topological semimetal. To date, only one polytype of GaGeTe is known with trigonal centrosymmetric structure ($\alpha$ phase, space group *R-3m*, No. 166). Here we show that as-grown samples of GaGeTe show traces of at least another polytype with hexagonal non-centrosymmetric structure ($\beta$ phase, space group *P6₃mc*, No. 186). Moreover, we suggest that another bulk hexagonal polytype ($\gamma$ phase, space group *P-3m1*, No. 164) could also be found near room conditions. Both $\alpha$ and $\beta$ polytypes have been identified and characterized by means of X-ray diffraction and Raman scattering measurements with the support of *ab initio* calculations. We provide the vibrational properties of both polytypes and show that the Raman spectrum reported for GaGeTe almost forty years ago and attributed to the $\alpha$ phase, was, in fact, that of the secondary $\beta$ phase. Additionally, we show that a Fermi resonance occurs in $\alpha$-GaGeTe under non-resonant excitation conditions, but not under resonant excitation conditions. Theoretical calculations show that bulk $\beta$-GaGeTe is a non-centrosymmetric weak topological semimetal with even smaller lattice thermal conductivity than centrosymmetric bulk $\alpha$-GaGeTe. In perspective, our work paves the way for the control and engineering of GaGeTe polytypes to design and implement complex van der Waals heterostructures formed by a combination of centrosymmetric and non-centrosymmetric layers of up to three different polytypes in a single material, suitable for a number of fundamental studies and technological applications.

1. Introduction

The study of topological semimetals (TSMs) is a hot topic in Condensed Matter Science. Many research efforts have been recently focused on the search for TSMs, both of Dirac and Weyl types, to find Dirac and Weyl fermions almost 100 years after their theoretical predictions [1]. In fact, Dirac TSMs can be driven either into a Weyl TSM or a topological insulator by symmetry breaking or by increasing spin-orbit coupling, respectively. TSMs have unusual electronic properties, such as extremely high carrier mobility, chiral effects, negative magnetoresistance, anomalous Hall effect, and low lattice thermal conductivity. These extraordinary effects promise to be useful for many devices and applications, like optical detectors, catalysis, spintronics, valleytronics, straintronics, and highly efficient thermoelectrics [2].

Since the discovery of the Dirac TSM $Cd_3As_2$, a three-dimensional analog to graphene [2], the research work on TSMs has greatly increased. Transition metal dichalcogenides, as well as Ta and Nb pnictogenides, have been thoroughly studied in the last decade and at present a plethora of new TSMs are under examination. In fact, non-centrosymmetric topological materials have attracted the most recent interest since they show non-trivial phenomena that could complement centrosymmetric variants in topological electronic devices. In particular, symmetry breaking allows many exotic properties to emerge in topological materials, such as the ferroelectricity and pyroelectricity [3], surface-dependent topological electronic states [4], as well as non-linear Hall effect and non-linear photocurrent [5-8].

Among TSMs, one of the less studied compounds is GaGeTe, which has been recently cataloged as a basic $Z_2$ TSM with very interesting properties [9]. GaGeTe is a layered material with a very small bandgap and huge carrier mobility that is composed of a germanene-like sublayer embedded between two GaTe sublayers [10,11] of InSe or GaSe-type [12]. Therefore, GaGeTe is a van der Waals (vdW) heterostructure that is chemically stable in air, water, and NaOH at room conditions. Additionally, it has been recently predicted to be both dynamically and thermodynamically stable at high temperatures [13] and it can be easily exfoliated, being very promising for 2D electronic and optoelectronic applications.

GaGeTe was synthesized more than 40 years ago [10] and soon after its structural [11] and vibrational [14] properties were investigated at room conditions. To our knowledge, only a single polymorph of this compound is known (hereafter, α phase). It crystallizes in a trigonal centrosymmetric structure (space group (s.g.) *R-3m*, No. 166) whose lattice parameters in a hexagonal setting are: *a* = 4.048 Å and *c* = 34.731 Å. α-GaGeTe is composed by three GaGeTe monolayers stacked along the *c* axis of the hexagonal unit cell (**Fig. 1**). Each layer of GaGeTe has *P-3m1* symmetry and consists of six atomic planes Te-Ga-Ge-Ge-Ga-Te perpendicular to the hexagonal c-axis (see **Fig. 1**), being the intralayer forces between atoms of covalent-type and the force between the layers of vdW type. In this layered structure, Ga and Ge atoms show tetrahedral coordination, while Te atoms are threefold coordinated, as expected from the covalent and vdW forces present in the crystalline structure.

From the theoretical point of view, initial studies suggested that bulk α-GaGeTe is a 3D strong topological insulator **[15]**. Moreover, monolayer and few-layer GaGeTe were predicted as promising 2D materials for electronic and optoelectronic applications, including solar cells and field-effect transistors, due to their high thermal stability, high electron mobility, and control of a wide tunable bandgap with the control of the number of monolayers or the compressive or tensile strain applied to the layers **[13]**. On top of that, theoretical calculations in monolayer α-GaGeTe predicted interesting spintronic properties that can be controlled by the modulation of an applied electric field due to the Rashba-type spin-orbit coupling **[16]**. These features would allow turning an indirect bandgap into a direct one, making monolayers of α-GaGeTe promising candidates in 2D nanoelectronics and spintronics **[16]**. More recently, *ab initio* calculations of the electronic properties of a Germanene/α-GaGeTe heterostructure have shown the ability to tune the bandgap via an applied voltage or strain, thus making possible a Germanene-based field-effect transistor **[17]**.

From the experimental point of view, α-GaGeTe has been studied for its possible use as a thermoelectric material **[18,19]**, finding that its relatively low power factor increases with temperature, making it a promising material for high-temperature applications. More recently, α-GaGeTe has been shown as a promising candidate for 2D transistors and photodetectors with ON/OFF current ratios larger than many other reported field-effect transistors based 2D materials and a photoresponsivity higher than $MoS_2$, graphene, and few-layer black phosphorus **[20]**. Furthermore, a short-wavelength infrared detector based on α-GaGeTe, with notable performance in terms of speed and bandwidth, has been recently demonstrated **[21]**.

Despite recent advances in the characterization of many properties of bulk and monolayer α-GaGeTe, several of their basic properties are still under discussion. For instance, the nature and value of the bandgap in bulk α-GaGeTe has been a recently resolved puzzle. Initial studies suggested that α-GaGeTe was a degenerate p-type semiconductor with a bandgap around 1.0 eV and low hole mobility around 40 $cm^2$ $V^{-1}$ $s^{-1}$ **[18,19]**. However, recent synchrotron-based high-resolution angle-resolved photoemission spectroscopy measurements, combined with band structure calculations, have shown that bulk α-GaGeTe has a very small indirect bandgap and features strong topological properties [9]. In fact, it has been proposed to be the first observed basic $Z_2$ TSM with three types of charge carriers: bulk electrons and holes as well as surface electrons. This discovery confirms the fundamental importance of α-GaGeTe for the understanding and applications of topological materials.

Given that electron-phonon interactions are crucial in TSMs, recent studies of TSMs have remarked the necessity of the study of the vibrational properties of these compounds, in order to understand their transport properties [22,23]. In particular, anomalous profiles of Raman-active modes, like Fano resonances [24-26], anomalous angle-dependence of Raman-active modes [27], double resonance Raman modes [28], and quasielastic scattering [29] have been observed in Raman scattering (RS) measurements. Moreover, RS measurements, naturally sensitive to the crystal symmetry, are very important to reveal the inversion symmetry breaking, since the non-centrosymmetry is a prerequisite for realizing Weyl fermions in non-magnetic materials [30]. In this context, very little is known regarding the vibrational properties of bulk α-GaGeTe. To the best of our knowledge, only one Raman study of GaGeTe at 77 and 297 K has been performed [14]. In that work, six Raman modes were observed for this TSM and attributed to five out of six Raman-active modes and one IR-active mode of the *R-3m* structure; however, several observed Raman bands could not be explained and the observed Raman modes still have not been discussed in the light of modern *ab initio* calculations. Additionally, the phonon dispersion curves and phonon density of states of bulk α-GaGeTe are still unknown. Curiously, the vibrational properties of the monolayer of α-GaGeTe (with s.g. *P-3m1*) have been recently studied and its lattice thermal conductivity was found to be comparable to that of MoS$_2$ and much lower than that of graphene [31], which promises a use in very efficient thermoelectric materials.

In this work, we have studied the RS spectra of bulk GaGeTe at room temperature by exciting our samples with two different laser lines (on- and off-resonance). Two different RS spectra have been mainly observed in bulk samples. In order to clearly establish which one of these is the actual RS spectra of α-GaGeTe, we have performed *ab initio* calculations that are compared to our RS measurements. As a result, we show that the previously reported Raman modes [14] do not correspond to α-GaGeTe, but to another polytype (β-GaGeTe) that is a secondary phase that coexists with α-GaGeTe. The new β polytype has non-centrosymmetric hexagonal structure (s.g. *P6$_3$mc*, No. 186). Moreover, we suggest the possible existence of a third polytype of bulk GaGeTe (γ-GaGeTe) with a centrosymmetric hexagonal structure (s.g. *P-3m1*, No. 164). The existence of β-GaGeTe has been confirmed not only by RS measurements but also by powder X-ray diffraction (XRD) measurements at ambient conditions. *Ab initio* calculations not only confirm that β-GaGeTe is energetically competitive with α-GaGeTe but they also show a nice match of the simulated structural and vibrational properties with those experimentally observed. Therefore, we provide here a complete structural and vibrational characterization of both α-GaGeTe and β-GaGeTe. In particular, we comment on the existence of a Fermi resonance in the mid-frequency modes of α-GaGeTe, but not in β-GaGeTe. The Fermi resonance in the RS spectrum of α-GaGeTe is clearly observed under non-resonant excitation conditions, but it is not observed under resonant excitation conditions. To our knowledge, this phenomenon has not been observed in solids before. Moreover, we confirm that bulk β-GaGeTe is a non-centrosymmetric weak TSM that exhibits a smaller lattice thermal conductivity than bulk α-GaGeTe at room conditions.

We believe that our work, with its in-depth experimental and theoretical analysis of both α and β polytypes, may stimulate further studies to control and engineer the different GaGeTe polytypes for a number of fundamental studies and technological applications, in particular those of vdW heterostructures implemented by stacking different polytypes of a single material.

## 2. Methods

### A. Experimental details

GaGeTe crystals grown using a modified Bridgman method as described in **Ref**. **18** and used in **Refs. 18 and 19** have been used in this work. Powder XRD measurements of as-grown samples have been performed with Cu $K_{\alpha 1}$ radiation on a Rigaku Ultima IV diffractometer to verify the crystalline structure of the samples. The structural analysis has been performed using the Rietveld method. For α-GaGeTe, we used as starting model the structure determined by Fenske *et al*. **[11]** and for β-GaGeTe, we used the structure determined from DFT calculations in this work.

Room-temperature unpolarized and polarized RS measurements on bulk GaGeTe crystals were performed with a Horiba Jobin Yvon LabRAM HR UV spectrometer equiped with a thermoelectrically cooled multichannel CCD detector. Raman signals were excited with both the 532-nm (green) line of a solid state laser and the 632.8-nm (red) line of a HeNe laser. For the green line, common edge filters allowing measurements above 30-50 cm$^{-1}$ were used. For the red line, an ultra-low frequency (ULF) filter composed of three Volume Bragg Grating notch filters allowed measurements down to 10 cm$^{-1}$. In all measurements, a spectral resolution smaller than 3 cm$^{-1}$ is obtained. RS measurements at high pressures were performed by inserting small 100x100x20 μm$^3$ samples into a diamond anvil cell and using a methanol-ethanol mixture as a pressure-transmitting medium and a ruby chip as pressure sensor [32].

### B. *Ab initio* calculation details

*Ab initio* total-energy calculations at 0 K for the *R-3m, P6$_3$mc, and P-3m1* phases of GaGeTe were performed within the framework of density functional theory (DFT) [33] with the Vienna Ab-initio Simulation Package (VASP) [34], using the pseudopotential method and the projector augmented waves (PAW) scheme [35,36]. The valence electron configurations of Ga, Ge, and Te atoms adopted are 3d$^{10}$4s$^2$4p$^1$, 3d$^{10}$4s$^2$4p$^2$, and 5s$^2$5p$^4$, respectively. In this work, the generalized gradient approximation (GGA) with the Perdew-Burke-Ernzerhof for solids (PBEsol) parametrization [37] and the Perdew-Burke-Ernzerhof [38] parametrization including the dispersion corrections of Grimme (PBE-D3) [39] were used for the exchange and correlation energy. A dense Monkhorst–Pack grid [40] of special k-points (6×6×6 for *R-3m*, 20x20x4 for *P6$_3$mc*, and 8x8x4 for *P-3m1* phases, respectively) along the Brillouin zone (BZ) and a plane-wave basis set with energy cutoffs of 540 eV were used. All atomic degrees of freedom, including lattice constants and atomic parameters, were fully relaxed with self-consistent convergence criteria of 0.004 eV/Å and 10$^{-6}$ eV for the atomic forces and the total energy, respectively.

Lattice-dynamical properties with VASP were obtained for the Γ-point of the BZ using the direct-force constant approach within the PBEsol and PBE-D3 prescription in order to compute the atomic forces. Additionally, the PHONON code [41,42] was used to simulate the RS spectrum of the different phases. This method involves the construction of a dynamical matrix at the Γ point of the BZ. Separate calculations of the atomic forces are needed and performed by small independent displacements of atoms from the equilibrium configuration within the primitive cell, whose number depends on the crystal symmetry. Highly converged results on forces are required for the calculation of the dynamical matrix [42]. The subsequent diagonalization of the dynamical matrix provides the frequencies of the normal modes. Moreover, these calculations allow the identification of the symmetry and eigenvectors of the vibrational modes in each structure at the Γ point. To obtain the phonon dispersion curves along high-symmetry directions of the BZ and the one-phonon density of states, we performed similar calculations using appropriate supercells, which allow the phonon dispersion at k-points to be obtained commensurate with the supercell size [42]. The J-ICE software [43] was used to plot the atomic vibrations of bulk α-GaGeTe and bulk β-GaGeTe given in the Electronic Supplementary Information (ESI). The two-phonon (or joint phonon) density of states (DOS) was calculated for bulk α-GaGeTe and bulk β-GaGeTe with the electronic structure ABINIT **[44,45]** using the same correction considered for the structural relaxation. The search of competitive phases was performed by using complementary approaches, on one side, we used symmetry arguments to build possible phases and on the other hand, we use a full structural search method. The Minima Hopping Method was adopted, following the original implementation **[46]** and with parameters similar to those used in **[47]**.

Lattice thermal conductivity and phonon lifetimes were calculated by employing the PHONO3PY code **[48]** and VASP is used as the calculator for the third-order (anharmonic) interatomic force constants. A 2 × 2 × 2 supercell is used, with a q mesh of 20 × 20 × 10 for both bulk α-GaGeTe and bulk β-GaGeTe with the tetrahedron method to perform the integration for the phonon lifetime calculation. The lattice thermal conductivity and phonon lifetimes were computed with the single-mode relaxation-time approximation, to solve the Boltzmann transport equations.

Finally, the topological properties of bulk α-GaGeTe and bulk β-GaGeTe were analysed by using an online tool which predicts the topological classification of materials with machine learning. the tool is based on gradient boosted trees trained with the ab-initio results from the topological quantum chemistry database **[49]**.

3. Results

A. Structural and vibrational properties of bulk GaGeTe polytypes

As previously commented, the centrosymmetric trigonal *R-3m* structure of bulk α-GaGeTe was established forty years ago **[10,11],** and soon after its vibrational properties were reported and discussed in relation to the *R-3m* structure **[14]**. In that work, it was reported that the RS spectrum of GaGeTe was different under resonant and non-resonant excitation conditions and that the RS spectra of GaGeTe showed more Raman modes than expected for the *R-3m* phase. Consequently, some modes were tentatively attributed to IR-active modes or second-order modes of the *R-3m* structure.

**Figure 2** shows the unpolarized and polarized (parallel and crossed) backscattering RS spectra of GaGeTe at room conditions when excited under non-resonant conditions with the 632.8 nm (1.98 eV) red laser line and under resonant conditions with the 532 nm (2.33 eV) green laser line. Unexpectedly, we observe two types of RS spectra: i) The dominant RS spectra of the as-grown sample (**Fig. 2a**) and ii) the RS spectra obtained in a few places of the same sample (**Fig. 2b**). Curiously, the RS spectra of **Fig. 2b** is similar to that reported almost 40 years ago by López-

Cruz *et al*. [14]. Eventually, both RS spectra could be simultaneously measured (see **Fig. S1** in ESI).

The RS spectra of **Figs. 2a and 2b** have notable similarities, as both are dominated by narrow peaks at comparable frequencies, especially at high frequencies (note that the absolute frequencies of the different modes are slightly different in both RS spectra). The largest difference is observed between the low-frequency A-type modes near 79 cm$^{-1}$ (**Fig. 2a**) and 66 cm$^{-1}$ (**Fig. 2b**). The similar high-frequency and different low-frequency modes clearly indicate that there must be at least two different polytypes of GaGeTe in as-grown samples. In fact, the possible existence of GaGeTe polytypism is no surprise since this compound is a vdW heterostructure of GaSe-type and several polytypes of GaSe, InSe, and GaS, generated by different piling sequences of the same layers along the hexagonal *c* axis, are known **[50,51]**.

In order to verify which of the two RS spectra really correspond to α-GaGeTe, we have performed theoretical first-principles simulations of the lattice dynamics of α-GaGeTe. According to group theory, α-GaGeTe with trigonal *R-3m* structure and 3 atoms per unit cell located at *6c* Wyckoff sites must exhibit 18 vibrational modes at the BZ center, Γ, corresponding to 10 optical modes (6 Raman-active modes, 3 $A_{1g}$ + 3 $E_g$, 4 infrared (IR)-active modes, 2 $A_{2u}$ + 2 $E_u$) and 3 acoustic modes, $A_{2u}$ + $E_u$ [52]. The ten optical vibrational modes at the BZ center will be noted hereafter with a superindex in order of increasing frequency (**Table 1**). The four lowest modes at Γ, corresponding to $E_g^1$, $E_u^1$, $A_{1g}^1$, and $A_{2u}^1$ modes in order of increasing frequency, correspond to shear (or transversal, E) and longitudinal (A) vibrations between neighbor Ge and GaTe sublayers that can be both intra-layer and inter-layer (see further details in **Figs. S2 and S3** in ESI). On the other hand, the modes at higher frequencies ($E_g^2$, $E_u^2$, $A_{1g}^2$, $A_{2u}^2$, $E_g^3$, $A_{1g}^3$) correspond to Ga-Te, Ga-Ge, and Ge-Ge intra-layer vibrations (see further details in **Figs. S4, S5 and S6** in ESI). For completeness, the phonon dispersion curves and the total and partial (atom projected) one-phonon density of states of α-GaGeTe are given in **Fig. S7** in ESI. As expected, the phonon dispersion curves of α-GaGeTe show that the *R-3m* structure is dynamically stable at room conditions since there are no imaginary frequencies along the BZ.

Notably, the number and frequencies of the Raman-active modes of α-GaGeTe are in better agreement with RS spectra of **Fig. 2a** than with those of **Fig. 2b** (see also **Table 1** for a comparison of the theoretical frequencies of α-GaGeTe and the experimental frequencies in **Fig. 2a** and in **Ref. 8**). Moreover, the theoretically simulated RS spectrum of α-GaGeTe at room conditions (**Fig. S8**) is similar to that of **Fig. 2a**. This result is consistent with the dominance of the RS spectra of **Fig. 2a** in as-grown samples that mainly correspond to the *R-3m* structure, as identified by the

Rietveld refinement of powder XRD measurements performed in the as-grown samples at room conditions (see **Fig. 3a** and **Table 2**). Note that our experimental and theoretical structural values for the *R-3m* phase compare nicely with those already reported **[10,11,18,19]**. Consequently, we can safely attribute the RS spectra of **Fig. 2a** to α-GaGeTe with *R-3m* structure and conclude that the RS spectra reported in **Fig. 2b** and in **Ref. 14** must correspond to a different polytype of GaGeTe.

For completeness, we provide in **Table 1** the different symmetries and wavenumbers of the experimentally observed modes of bulk α-GaGeTe in **Fig. 2a**, as obtained from polarized RS measurements and analyzed on the basis of theoretical *ab initio* calculations. The A or E symmetry of the different Raman-active modes in **Fig. 2a** has been determined thanks to polarized measurements with both exciting lasers, since A modes are excluded in rhombohedral and hexagonal layered materials in RS measurements performed in backscattering configuration for cross polarization. When excited with the red laser, the unpolarized RS spectrum of bulk α-GaGeTe is dominated by the mode near 280 cm$^{-1}$ ($E_g^3$ mode). Additionally, two narrow bands at 41 ($E_g^1$ mode) and 79 cm$^{-1}$ ($A_{1g}^1$ mode) and two broad bands around 266 ($A_{2u}^2$ mode) and 296 cm$^{-1}$ ($A_{1g}^3$ mode) are observed. Finally, a number of broad bands located near the expected positions for the two mid-frequency ($E_g^2$ and $A_{1g}^2$) modes are observed. When excited with the green laser, the unpolarized RS spectrum is clearly dominated by the $A_g^1$ mode. Additionally, four narrow bands are observed: the $E_g^3$ and $A_{1g}^3$ modes and two additional bands at 177 and 206 cm$^{-1}$ that we have attributed to the $E_g^2$ and $A_{1g}^2$ modes. These two last modes were not observed under non-resonant excitation conditions (we will comment later on this subject). Unfortunately, the low-frequency $E_g^1$ mode could not be observed under green excitation, likely due to both its weak signal and the overlap with the strong Rayleigh scattering (it could not be efficiently eliminated by our edge filter for the green laser).

Once clarified the actual RS spectrum of bulk α-GaGeTe, the next question is: Which is the polytype that shows the RS spectra of **Fig. 2b** and **Ref. 14**? To answer that question, a join effort of symmetry arguments and structural search **[46,47]** has been undertaken. We will show that, since α-GaGeTe is a complex polytype with three monolayers stacked along the c axis, like γ-InSe and γ-GaSe **[50,51]**, the answer is given by looking for possible polytypes of GaGeTe with one or two monolayers of GaGeTe stacked along the hexagonal c axis.

The simplest polytype one can consider (hereafter named GaGeTe-mono) is composed of α-GaGeTe monolayers piled on top of each other; i.e. without the shift of neighbor layers along the hexagonal a-b plane observed in α-GaGeTe (**Fig. 1a**). This polytype belongs to s.g. *P-3m1* and contains only one monolayer of GaGeTe along the hexagonal c axis (see **Fig. S9**). *Ab initio* simulations of the vibrational properties of this polytype, whose structure is summarize in **Table S1** in ESI, show that its vibrational modes are similar to those of α-GaGeTe; i.e. 18 vibrational modes at Γ corresponding to 10 optical modes (6 Raman-active modes, 3 $A_{1g}$ + 3 $E_g$, 4 infrared (IR)-active modes, 2 $A_{2u}$ + 2 $E_u$) and 3 acoustic modes, $A_{2u}$ + $E_u$ **[52]**. Additionally, the frequencies of this polytype are very similar to those of α-GaGeTe (see **Table S2**); however, our calculations show that this polytype is dynamically unstable since it has imaginary frequencies along the BZ (see **Fig. S9**). Moreover, this polytype is not energetically competitive with respect to α-GaGeTe as seen in **Fig. 4**. These theoretical results are expected since this polytype shows a very low packing efficiency of the layers along the hexagonal *c* axis that gives rise to very strong repulsion forces between Te atoms. These forces, caused by the repulsion of the lone electron pairs of Te atoms, are very strong because Te atoms of adjacent layers are one on top of the other. This atomic disposition is in contrast to that found in α-GaGeTe, in which the different layers are well packed along the *c* axis due to the shift of the layers along the a-b plane. It can be seen that short and long Ga-Te bonds viewed in **Fig. 1** form a zigzag structure and there is a coupling of the zigzag structure of neighbor layers, thus reducing the repulsion between neighbor Te atoms.

Two additional polytypes of GaGeTe have been found to be relevant for this study. One of them, hereafter named γ-GaGeTe, is also a polytype containing one layer along the hexagonal *c* axis and with s.g. *P-3m1* (see **Fig. 1**). This polytype is formed by monolayers that are different from those observed in α-GaGeTe. The monolayers of γ-GaGeTe show a 180° rotated germanene sublayer with respect to the monolayers of α-GaGeTe. Consequently, top and bottom Te atoms in the γ-GaGeTe monolayer are not aligned the hexagonal *c* axis. This arrangement of top and bottom Te atoms in the γ-GaGeTe monolayer allows an efficient zig-zag packing of layers in bulk γ-GaGeTe, similar to bulk α-GaGeTe, but with the hexagonal *c* axis containing only one monolayer instead of the three found in bulk α-GaGeTe. As expected, bulk γ-GaGeTe is energetically competitive with bulk α-GaGeTe (see **Fig. 4**). Moreover, bulk γ-GaGeTe is dynamically stable at room conditions according to the theoretical phonon dispersion curves and one-phonon density of states of bulk γ-GaGeTe at room conditions (**Fig. S10**). Therefore, we can conclude that this polymorph might be observed in as-grown GaGeTe samples at room conditions.

For the sake of comparison with experimental measurements, we provide the theoretical structural and vibrational parameters of bulk γ-GaGeTe at 0 GPa and 0 K in **Tables S3 and S4,** respectively. *Ab initio* simulations suggest that bulk γ-GaGeTe gives similar vibrational modes to those of bulk α-GaGeTe; i.e. 18 vibrational modes at the BZ center corresponding to 10 optical modes (6 Raman-active modes, 3 $A_{1g}$ + 3 $E_g$, 4 infrared (IR)-active modes, 2 $A_{2u}$ + 2 $E_u$) and 3 acoustic modes, $A_{2u}$ + $E_u$ **[52]**. As observed in **Table S4**, the Raman- and IR-active frequencies are very similar to those of bulk α-GaGeTe; therefore, this polymorph is not able to explain the RS spectra of **Fig. 2b**. On the other hand, we must note that this polymorph could explain the RS spectra of **Fig. 2a**; however, the simulation of the unpolarized non-resonant RS spectrum of bulk γ-GaGeTe (**Fig. S8**) is completely different to that of bulk α-GaGeTe. In bulk γ-GaGeTe, the RS spectrum is dominated by A-type modes, while in bulk α-GaGeTe it is dominated by E-type modes. Therefore, non-resonant RS spectra of **Fig. 2a** are consistent with bulk α-GaGeTe and not with bulk γ-GaGeTe. Moreover, both polymorphs can be distinguished by XRD measurements. In fact, we could not do a Rietveld refinement of the XRD patterns of the majority phase with the γ-GaGeTe structure since neither the number nor the intensity of the peaks match. Consequently, we can exclude that either RS spectra of **Fig. 2a** or of **Fig. 2b** correspond to γ-GaGeTe. At present, we have not found this polytype in as-grown GaGeTe samples.

The other possible polymorph that we have considered contains two layers along the hexagonal *c* axis (see **Fig. 1**) and has a non-centrosymmetric structure (s.g. *P6₃mc*, No. 186). This polytype (hereafter named β-GaGeTe) is formed by monolayers identical to those of α-GaGeTe; however, the second monolayer along the *c* axis is the specular image of the monolayer of α-GaGeTe slightly shifted along the a-b plane. In this arrangement, the packing of the two monolayers in the hexagonal unit cell is not perfect because the zig-zag sequence of the neighbor layers is not the same, but the repulsion between the monolayers is relatively weak because Te atoms of adjacent layers are not aligned along the *c* axis. As expected, β-GaGeTe is energetically competitive with α-GaGeTe and γ-GaGeTe at room conditions (see **Fig. 4**). Moreover, β-GaGeTe is dynamically stable at room conditions, according to the theoretical phonon dispersion curves and one-phonon density of states of β-GaGeTe at room conditions (**Fig. S11**). Therefore, we conclude that this polytype might also be observed in as-grown GaGeTe samples.

Again, for the sake of comparison with experimental measurements, we provide the theoretical vibrational frequencies and structural parameters of bulk β-GaGeTe at 0 GPa and 0 K in **Tables 3 and 4**, respectively. *Ab initio* simulations of the vibrational properties of bulk β-GaGeTe show that its vibrational modes are completely different to those of bulk α-GaGeTe. Group theoretical considerations of the *P6₃mc* structure, with 2 atoms in *2a* Wyckoff sites and 4 atoms in *2b* sites, yield 36 normal modes of vibration at Γ, whose mechanical decomposition is **[52]**: $\Gamma_{36}$= 5 $A_1$(R,IR)

+ 6 B$_1$(S) + 5 E$_1$(R,IR) + 6 E$_2$(R) + A$_1$(ac) + E$_1$(ac), where R, IR, S, and ac correspond to Raman-active, IR-active, silent, and acoustic modes, respectively. Therefore, there are a total of 16 Raman-active modes that can be up to 26 if we consider the TO-LO splitting of polar A$_1$ and E$_1$ modes that are both Raman and IR-active. The atomic vibrations of all the Raman- and IR-active modes of bulk β-GaGeTe have been plotted in **Figs. S12 to S22** in ESI. We have to note that most of the vibrational modes of bulk β-GaGeTe are related to bulk α-GaGeTe. In fact, there is a double number of vibrational modes in bulk β-GaGeTe than in bulk α-GaGeTe due to the double number of atoms in the primitive unit cell. In this way, the modes of bulk β-GaGeTe come in pairs that have their single parent vibrational mode in bulk α-GaGeTe.

Before commenting on the vibrational modes of bulk β-GaGeTe appearing in the RS spectra of **Fig. 2b**, we want to stress that these RS spectra exhibit two broad bands around 127 and 143 cm$^{-1}$ that were also reported by López-Cruz *et al*. **[14]**. These authors suggested that these two bands could be attributed to the second-order scattering of bulk α-GaGeTe, but they also commented that this hypothesis was not confirmed by their temperature-dependent measurements. It has been recently demonstrated that these two broad bands, appearing in the RS spectrum of many tellurides (including GaGeTe), correspond to the two most intense Raman-active modes of trigonal Te **[53]**. This element has three main Raman-active modes (one A$_1$ and two E modes). The Raman-active modes of Te appear in a number of tellurides because of the formation of nano- or micro-clusters of Te either present in as-grown samples or induced by sample degradation **[53,54]**. The RS spectrum of some regions of GaGeTe samples only show these two broad modes, as already reported **[53]**, thus suggesting that there are regions of as-grown GaGeTe samples with segregated nano- or micro-clusters of Te. Further information regarding the Raman modes of Te precipitates in a number of tellurides can be found in **Ref. 53**. It is important to comment that in **Ref. 53** it was assumed that the normal RS spectrum of bulk α-GaGeTe was the one reported by López-Cruz *et al*. because at that time the existence of different polytypes of GaGeTe was not taken into account. In this work we take an opposite view, inferring the existence of different GaGeTe polymorphs from experimental measurements and theoretical considerations.

Now that we have clarified the Te-related origin of some bands appearing in **Fig. 2b** and also in the RS spectra reported in **Ref. [14]**, we can comment on the rest of the modes present in **Fig. 2b** that we will show that correspond to bulk β-GaGeTe. In this context, the number and frequencies of the Raman-active modes observed in the RS spectra of **Fig. 2b** show a good agreement with the theoretically predicted transversal optic (TO) modes of bulk β-GaGeTe (see bottom marks in **Fig. 2b** and **Table 3**). Since no experimental LO mode seems to be observed, we have not performed theoretical calculations of longitudinal optic (LO) modes. The lack of observation of LO modes suggests a very small LO-TO splitting and a very small ionicity of this compound. Unfortunately, we have not been able to theoretically simulate the RS spectrum of β-GaGeTe at room conditions and compare it to experimental data **[55]**.

Similarly to α-GaGeTe, the A or E symmetry of the different Raman-active modes of β-GaGeTe has been determined thanks to polarized measurements with both exciting lasers (**Fig. 2b**) and our results are coincident to those reported in **Ref. 14**. As shown in **Fig. 2b** and **Table 3**, most observed E-type modes seem to be $E_1$ (TO) modes, while $E_2$ modes are either not observed (some $E_1$ and $E_2$ modes are almost coincident in frequency and may overlap) or are very weak ($E_2$ modes of very small frequency are not observed and only the $E_2^2$ mode seems to have been observed under resonance conditions close to the $E_1^1$ mode). Therefore, we can conclude that the RS spectra of **Fig. 2b** and of **Ref. 14** can be explained as the vibrational modes of bulk β-GaGeTe. Importantly, the assignment of the Raman modes of **Fig. 2b** to bulk β-GaGeTe allows us to explain the extra modes in **Fig. 2b** with respect to α-GaGeTe as being due to Raman-active modes and not to IR-active modes of α-GaGeTe as assumed in **Ref. 14**. Note that a number of Raman-active modes of bulk β-GaGeTe come from IR-active modes of bulk α-GaGeTe since bulk β-GaGeTe is a non-centrosymmetric structure with double number of vibrational modes at Γ than bulk α-GaGeTe, as already commented.

Our claim for the new polytype β-GaGeTe as being responsible for the RS spectra of **Fig. 2b** is confirmed by powder XRD measurements performed in as-grown samples at room conditions. In some samples, we have measured XRD patterns showing only peaks of bulk α-GaGeTe (see **Fig. 3a**); however, in other simples, we have measured XRD patterns with additional peaks (see **Fig. 3b**). The XRD patterns with additional peaks can be very well explained assuming the coexistence of α-GaGeTe and β-GaGeTe. In particular, according to peak intensities, **Fig. 3b** corresponds to a mixture of 70% α-GaGeTe and 30% β-GaGeTe. The experimental structural parameters of bulk β-GaGeTe obtained by the Rietveld refinement of XRD patterns in **Fig. 3b** are given in **Table 4** and show a good agreement with our theoretical data.

It must be stressed that despite most modes of bulk β-GaGeTe are similar in frequency, intensity, and polarization to those of bulk α-GaGeTe, they are shifted in frequency, as already pointed out. This is consistent with what could be expected for a polytype whose main difference is likely a change in the piling of the layers along the hexagonal *c* axis. In particular, it can be observed that the low-frequency (high-frequency) modes of bulk β-GaGeTe are shifted to low (high) values with respect to bulk α-GaGeTe. The smaller frequencies in the low-frequency region in bulk β-GaGeTe are indicative of a smaller interlayer interaction in this polytype than in bulk α-GaGeTe. This result is consistent with the higher repulsion between adjacent Te atoms in bulk β-GaGeTe than in bulk α-GaGeTe. A similar result has been already found in GaSe polytypes [51].

We comment now in detail the RS spectrum of bulk β-GaGeTe (**Fig. 2b**) on the light of the theoretical frequencies summarized in **Table 3**. When excited with the red laser, this RS spectrum is dominated by a band near 286 cm$^{-1}$ accompanied by two weak peaks at both sides of the main band at 263 and 297 cm$^{-1}$, as in bulk α-GaGeTe. These three modes correspond to the $E_1^5$, $A_1^4$, and $A_1^5$ modes of bulk β-GaGeTe, respectively. Three additional narrow bands are also observed at 35, 59, and 66 cm$^{-1}$, corresponding to the $E_1^1$, $E_1^2$, and $A_1^1$ modes of bulk β-GaGeTe, respectively. We have to stress that the two lowest-frequency modes (below 64 cm$^{-1}$) of bulk β-GaGeTe in **Fig. 2b** were not previously reported because of the strong Rayleigh scattering in **Ref. 14**. In addition, there are two mid-frequency modes between 170 and 210 cm$^{-1}$; i.e. in the region where some modes of bulk β-GaGeTe are also expected. These modes are poorly observed when excited with the red laser, with only one narrow peak around 167 cm$^{-1}$ being well observed. This mode has been tentatively attributed to the $A_1^2$ mode, despite its theoretically predicted frequency in bulk β-GaGeTe is much smaller (ca. 139 cm$^{-1}$).

A RS spectrum of bulk β-GaGeTe similar to that reported by López-Cruz *et al*. [14] was obtained under resonant conditions with the green laser. In this case, the RS spectrum is dominated by very strong modes near 66 cm$^{-1}$ ($A_1^1$ mode) and 263, 286, and 297 cm$^{-1}$ ($A_1^4$, $E_1^5$, and $A_1^5$ modes). Additionally, two narrow lines near 167 and 180 cm$^{-1}$ were found in the mid-frequency region. We have tentatively attributed these two lines to the $A_1^2$ and $E_1^3$ modes of bulk β-GaGeTe. We must also note that broad bands are observed between 180 and 215 cm$^{-1}$ in bulk β-GaGeTe both under resonant and non-resonant conditions, so we speculate that these broad bands could be related to the mid-frequency $E_1^4$ and $A_1^3$ modes of bulk β-GaGeTe; i.e. at similar frequencies to those in bulk α-GaGeTe under resonant conditions.

Once that most of the vibrational modes found in **Figs. 2a** and **2b** have been explained, we want to comment on the absence of certain modes in the RS spectra of both polytypes. It is noteworthy that we have not clearly detected the mid-frequency $A_{1g}^2$ and $E_g^2$ modes in bulk α-GaGeTe nor the mid-frequency $A_1^3$ and $E_1^4$ modes in bulk β-GaGeTe when measuring under non-resonant excitation conditions, but that $A_{1g}^2$ and $E_g^2$ modes in bulk α-GaGeTe have been measured under resonant excitation conditions. More specifically, very broad bands with different maxima have been measured in the mid-frequency region of bulk α-GaGeTe when measurements have been performed under non-resonant excitation conditions, while two narrow peaks were observed in that region when measurements have been performed under resonant excitation conditions.

Here we will show that the poor observation of the two mid-frequency Raman-active modes either in bulk $\alpha$-GaGeTe ($E_g^2$ and $A_{1g}^2$ modes) or in bulk $\beta$-GaGeTe ($E_1^4$ and $A_1^3$ modes) under non-resonant excitation conditions is likely due to the existence of a Fermi resonance in bulk $\alpha$-GaGeTe and to a strong anharmonic decay in bulk $\beta$-GaGeTe. To verify these mechanisms, we have followed the evolution of the two mid-frequency Raman-active modes of bulk $\alpha$-GaGeTe and $\beta$-GaGeTe as a function of pressure when excited with the red and green lasers; i.e. under non-resonant and resonant excitation conditions, respectively (see **Fig. 5**).

In bulk $\alpha$-GaGeTe, the two broad bands of the mid-frequency region can be followed under pressure when excited with the red laser (**Fig. 5a**). Above 5 GPa, a narrow band near 240 cm$^{-1}$ is observed, while a second narrow band can be observed at much higher pressures (above 8 GPa). On the other hand, the two narrow peaks in the mid-frequency region are observed when excited with the green laser at room pressure (**Fig. 5b**) and are nicely followed up to 2 GPa. Above this pressure, both peaks develop into intense broad peaks with adjacent bands and even into weak broad bands similar to those observed under non-resonant excitation conditions. In particular, the RS spectra of bulk $\alpha$-GaGeTe around 5.5 GPa excited with both the red and green lasers are very similar. This result indicates that resonant excitation conditions for RS measurements near room pressure with the green laser are no longer maintained above 2 GPa due to the shift of electronic levels with pressure. On increasing pressure and above 6 GPa, two narrow bands are observed in the RS spectra excited with the green laser in a similar way as in the one excited with the red laser. Since the pressure dependence of the experimental frequencies of the two modes measured with the green laser agrees with the theoretically expected behavior (**Fig. 5c**), it can be safely concluded that the bands observed at different pressures in this frequency range are clearly related to the $E_g^2$ and $A_{1g}^2$ modes **[56]**.

Now we will give an explanation of the features mentioned above in the mid-frequency region of the RS spectra in bulk $\alpha$-GaGeTe. It is well known that strong broadening of vibrational modes can occur due to anharmonic processes involving the decay of modes into sums and differences of two or more phonons, as evidenced in ZnO **[57]** and CuI **[58]**; however, the strong distortion of the two $E_g^2$ and $A_{1g}^2$ modes near 200 cm$^{-1}$ in bulk $\alpha$-GaGeTe at room pressure that shows broad bands with several relative maxima at low pressures and the development of two clear narrow modes above 5 GPa is consistent with the observation of a Fermi resonance at low pressures and the disappearance of the resonance at high pressures. Notably, similar pressure-induced de-activation of the Fermi resonance has been observed in several compounds, such as GaP, CuCl, CuBr, and CuI, and more recently $SnSb_2Te_4$ **[58-70]**. It must also be noted that the reverse process; i.e. pressure-induced activation of Fermi resonance, can lead to observation of Fermi resonance not present at room conditions, like in ice VII **[71]**.

The Fermi resonance is a very interesting phenomenon observed in Raman and IR spectroscopy, firstly observed in $CO_2$ in 1929 **[72]** and explained by Enrico Fermi in 1931 **[73]**. This phenomenon is very common in molecules and molecular crystals and it has been also observed in several crystalline solids, including graphene **[74]**. The simplest case occurs when the double excitation (first overtone) of a first-order phonon with frequency $\omega_1$ matches the energy of a second first-order phonon with frequency $\omega_2$ ($\omega_2 \approx 2\omega_1$). If both excitations have the same symmetry, they can mix or couple, thus originating an anharmonic decay of the second first-order phonon. As a consequence, two coupled excitations, $\omega_+$ and $\omega_-$, with different energies than those of the original coupled modes can be observed in the vibrational spectra. In general, the second first-order phonon looses its intensity, and the first overtone, usually of much smaller intensity than first-order phonons, can be partially observed since the intensity of both original modes is conserved **[75]**. The most complex case, typical of solids, consists of first-order modes that interact with a continuum of states that usually has a very high two-phonon (sum or difference) DOS (also known as the joint phonon DOS). The strong decay processes of the first-order phonon that is in resonance with an intense joint DOS leads to a change in frequency (renormalization of the phonon frequency), intensity, and width of the first-order phonon leading to the appearance of a three-mode aspect as that shown by the two bands in $\alpha$-GaGeTe at low pressures. As already mentioned, a similar phenomenon was earlier described in GaP, CuCl, CuBr, CuI, and also studied with increasing pressure in GaP, CuCl, CuBr, CuI, and $SnSb_2Te_4$ **[58-70]**.

Similarly to the cases of CuCl and CuBr **[66-69]**, the two broad bands with different maxima present in the mid-frequency region of bulk $\alpha$-GaGeTe are caused by a Fermi resonance that disappears with increasing pressure above 8.5 GPa. The disappearance of the Fermi resonance with increasing pressure is due to the stronger blueshift of the bare $E_g^2$ and $A_{1g}^2$ phonon frequencies with pressure than that of the two-phonon density of states resonating with them. It can be observed that the Fermi resonance in the RS spectra excited with the red laser disappears above 5.4 GPa for the high-frequency $A_{1g}^2$ mode, while for the $E_g^2$ mode it disappears at higher pressure. Surprisingly, two narrow bands corresponding to the experimental $E_g^2$ and $A_{1g}^2$ phonons are observed between room pressure and 2 GPa when excited with the green laser; i.e. Fermi resonance is not observed under resonant excitation conditions. Above 2 GPa, resonant excitation with the green laser disappears and RS spectra of higher pressure show the Fermi resonance as in the case of non-resonant excitation. We have to note that this phenomenon has not been observed at least in solids, to our knowledge, even in GaP, the most studied inorganic solid by means of RS measurements. In this context, we have to stress that GaP has indirect excitonic levels between 2.3 and 2.6 eV and direct excitonic levels around 2.78 eV at 300 K **[76,77]** and RS measurements in GaP have been extensively performed both under non-resonant and resonant excitation conditions **[59-63,65,77-79]**.

Support for the existence of a Fermi resonance in bulk α-GaGeTe is given by the comparison of the frequencies of the first-order phonons in this polytype and the bands of the two-phonon DOS (**Fig. S23** in ESI) at 0 GPa. In particular, the Fermi resonance of the $A_{1g}^2(\Gamma)$ phonon can be explained by: i) the almost exact coincidence of the frequency of this mode (197.9 cm$^{-1}$) with that of the difference of the $A_{1g}^3(\Gamma)$ and $A_{1g}^1(\Gamma)$ phonons (282.6 – 76.1 cm$^{-1}$) and ii) the strong two-phonon DOS of A-type near 200 cm$^{-1}$ caused by the strong one-phonon DOS of these last two A modes centered around 275 and 85 cm$^{-1}$, respectively (see **Fig. S7**). Note that sum and differences of two of these three modes could also lead to Fermi resonances of $A_{1g}^1(\Gamma)$ or $A_{1g}^3(\Gamma)$ phonons; however, these are prevented by the weak one-phonon DOS of the A-type mode between 195 and 215 cm$^{-1}$ (see **Fig. S7**). On the other hand, the Fermi resonance of the $E_g^2(\Gamma)$ phonon is not as easy to explain, since the frequency of this mode does not match with the $E_g^3(\Gamma) - E_g^1(\Gamma)$ difference and phonons of different parts of the BZ must be taken into account. In summary, we conclude that the observation of broad bands in the mid-frequency region of bulk α-GaGeTe in non-resonant RS measurements at room pressure is consistent with a Fermi resonance effect affecting both $A_{1g}^2(\Gamma)$ and $E_g^2(\Gamma)$ phonons and that Fermi resonance is not observed under resonant excitation conditions.

We will now explain the Raman modes observed in the mid-frequency region in bulk β-GaGeTe. Six Raman-active modes ($A_1^2$, $E_1^3$, $E_2^3$, $E_1^4$, $E_2^4$, $A_1^3$) are predicted in this region (**Table 3**). Leaving aside the two undetected $E_2$ modes, the $A_1^2$ and $E_1^3$ modes are clearly observed and followed under pressure. However, no bands related to the $E_1^4$ and $A_1^3$ modes could be followed under pressure excited either with the red laser (**Fig. 5d**) or the green laser (**Fig. 5e**), as observed by the pressure dependence of the experimental and theoretical frequencies measured in this mid-frequency region (**Fig. 5f**). Therefore, taking into account that our RS measurements on bulk β-GaGeTe with the green laser close to room pressure are of resonant type (as proved in **Ref. 14**) and in the light of the previous results on bulk α-GaGeTe for resonant and non-resonant excitation, we tentatively conclude that the lack of both Raman-active $E_1^4$ and $A_1^3$ modes in bulk β-GaGeTe cannot be due to a Fermi resonance, despite the two-phonon DOS of bulk β-GaGeTe being very similar to that of bulk α-GaGeTe (**Fig. S23** in ESI). Instead, we believe that these two modes cannot be detected because of their weak RS cross-section, due to a strong anharmonic decay of these two modes in bulk β-GaGeTe as a consequence of being in a region with a high density of states. We must note that the case of bulk β-GaGeTe compared to bulk α-GaGeTe is similar to that of zincblende-type CuI compared to isostructural CuBr and CuCl. While Fermi resonances were observed in CuCl and CuBr **[64-69]**, only a strong anharmonic decay was observed in CuI **[58]**.

### B. Lattice thermal conductivity of bulk α-GaGeTe and β-GaGeTe polytypes

It has been shown in the prototypical Dirac TSM $Cd_3As_2$ that the existence of soft optical phonon modes affects the lattice thermal conductivity (which ranges from 0.3 to 0.7 $Wm^{-1}K^{-1}$ at 300 K) **[80]**. The low-frequency optical phonon modes increase the available phase space of the phonon-phonon scattering of heat-carrying acoustic phonons. Consequently, this effect will cause low lattice thermal conductivity values for $Cd_3As_2$. Moreover, it has been shown that the interplay between the phonon-phonon Umklapp scattering rates and the soft optical phonon frequency explains the unusual non-monotonic temperature dependence of the lattice thermal conductivity of $Cd_3As_2$ **[80]**. Low-frequency optical phonon modes near Z and A are also present in the phonon dispersion curves of bulk α-GaGeTe (**Fig. S7**) and bulk β-GaGeTe (**Fig. S10**), respectively. These points are related to the *c*-axis direction of the hexagonal unit cell; i.e. perpendicular to the GaGeTe monolayers in the trigonal and hexagonal primitive unit cells of α-GaGeTe and β-GaGeTe, respectively.

Based on the third-order interatomic force constants, we have calculated the lattice thermal conductivity ($\kappa_L$) for bulk α-GaGeTe and β-GaGeTe as a function of temperature (**Fig. 6**). Due to the existence of low-frequency phonons it is not surprising to observe that the $\kappa_L$ of both polytypes are quite low. At room-temperature, $\kappa_L$ is 0.135 $Wm^{-1}K^{-1}$ along the two crystallographic x and y axis (in-plane), while along the layered z axis (out-of-plane), $\kappa_L$ lowers to 0.007 $Wm^{-1}K^{-1}$ for bulk α-GaGeTe. On the other hand, $\kappa_L$ is 0.109 $Wm^{-1}K^{-1}$ along the two crystallographic x and y axis (in-plane), while along the layered z axis (out-of-plane), $\kappa_L$ lowers to 0.006 $Wm^{-1}K^{-1}$ for bulk β-GaGeTe. These values compare to those just calculated for monolayer α-GaGeTe and can be compared to those recently reported for other topological materials, such as β-$As_2Te_3$ **[81]**. For the latter, a low-bandgap semiconductor with topological properties, the room-temperature $\kappa_L$ was calculated to be 0.098 Wm$^{-1}$K$^{-1}$ along the two crystallographic x and y axis (in-plane), while along the layered z axis (out-of-plane), $\kappa_L$ lowers to 0.023 Wm$^{-1}$K$^{-1}$ near room conditions. Therefore, we conclude that the thermoconductivity of $\beta$-GaGeTe is slightly smaller than that of $\alpha$-GaGeTe and that both compounds show even smaller values than in β-As$_2$Te$_3$ along the z axis. In this context, we must stress that calculations in these three materials show smaller values of lattice thermal conductivity than in most interesting thermoelectric compounds **[82]** and even smaller than those of Sb$_2$Te$_3$ and Bi$_2$Te$_3$, the most used thermoelectric materials at ambient conditions **[83,84]**.

The ultralow lattice thermal conductivity of both GaGeTe polytypes is not unexpected because it is well known that a low lattice thermal conductivity is usually associated to the low vibrational frequencies of heavy atoms **[85]**. In this context, the existence of very-low-frequency optical phonon modes is also observed in the phonon dispersion curves of the present calculations for both GaGeTe polytypes. Moreover, it is observed that both GaGeTe polytypes exhibit a monotonic decrease of $\kappa_L$ upon increasing temperature that follows roughly a T$^{-1}$ power law. Such a feature suggests that Umklapp scattering events dominate the thermal transport in this temperature range, as recently observed in β-As$_2$Te$_3$ **[81,86,87]**. In this sense, it has been also shown that the interplay between the phonon-phonon Umklapp scattering rates and the soft optical phonon frequency explains the unusual nonmonotonic temperature dependence of the lattice thermal conductivity of Cd$_3$As$_2$ **[80]**.

The small values of lattice thermal conductivity of both polytypes of GaGeTe can be also understood with the frequency-dependent phonon lifetimes at 300 K (see **Fig. S24**). Note that the short phonon lifetimes (below 2 ps) are much lower than those found for SnSe (from 0 to 30 ps) **[88]**. Moreover, our values of the phonon lifetimes are even lower than those found for ZrTe$_5$ **[89]**. The short lifetimes of the phonons indicate a strong scattering rate, which is the main reason for the ultralow lattice thermal conductivity **[90]**. Consequently, our calculations of phonon lifetimes allow us to explain the nature of the phonons which cause the low lattice thermal conductivity of bulk $\alpha$-GaGeTe and $\beta$-GaGeTe. We have also to note that both GaGeTe polytypes have large PDOS around 2 THz, thus reinforcing the role of these phonons in the ultralow lattice thermal conductivity. Additionally, a large PDOS is also observed in the mid-frequency region (around 5.2 THz, 173 cm$^{-1}$) in $\beta$-GaGeTe. Such a large PDOS is not observed in $\alpha$-GaGeTe and allows to explain why mid-frequency phonons of $\alpha$-GaGeTe show Fermi resonance while those of $\beta$-GaGeTe are not observed due to the strong decay of the phonons in this frequency region, as commented in the previous section.

**C. Electronic band structure of bulk α-GaGeTe and β-GaGeTe polytypes**

Due to the relevance of the electronic and topological properties of GaGeTe, we want to study the electronic band structure of both polytypes of GaGeTe from the theoretical point of view, since this point has raised considerable controversy in the last years **[9,13,15,17,18,21]**. In this context, a recent work has shown that experimental measurements agree more with GGA-PBE calculations including SOC interaction than with hybrid HSE06 calculations including SOC interaction that show a considerable overestimation of the bandgap **[9]**. Therefore, we have calculated the electronic band structure of bulk α-GaGeTe and β-GaGeTe at room conditions using GGA-PBEsol functionals including the SOC interaction (see **Fig. 7**).

The band structure of bulk α-GaGeTe is represented along the U-Γ-Z-F-Γ-L directions of the *R-3m* structure (note that the Z point is the same as the T point in **Refs. 9 and 15**). It can be observed that the valence band maximum (VBM) at room pressure is along the Z-F direction close to the Z point (hereafter Z' point). On the other hand, the conduction band minimum (CBM) is in the U-Γ direction close to the U point (hereafter U' point), although there is also a close minimum at the F-Γ direction close to the F point (hereafter F' point). Therefore, according to our calculations bulk α-GaGeTe is an indirect semiconductor (almost a semimetal) with small bandgaps: an indirect Z'-U' bandgap of 8 meV, another indirect Z'-Z bandgap of 54 meV, and a direct bandgap at the Z point of 120 meV.

On the other hand, the band structure of bulk β-GaGeTe is represented along the Γ-M-K-Γ-A-L-H-A directions of the *P6$_3$mc* structure, typical of wurzite-like materials. The valence band maximum (VBM) at room pressure is along the Γ-K direction close to the Γ point (hereafter Γ' point) although there is a close maximum along the Γ-M direction. The conduction band minimum (CBM) is at the Γ point, although there is also a close minimum near the M point. Therefore, according to our calculations bulk β-GaGeTe is also an indirect semiconductor (almost a semimetal) with small bandgaps: an indirect Γ'-Γ bandgap of 3 meV and a direct bandgap at the Γ point of 61 meV; i.e. bulk β-GaGeTe shows even smallest bandgaps than bulk α-GaGeTe.

Finally, we have calculated the parities of the states at the time-reversal invariant momentum (TRIM) points of the primitive rhombohedral cell of bulk α-GaGeTe and the primitive hexagonal cell of bulk β-GaGeTe at room conditions in order to check their topological properties by the calculation of the four topological Z$_2$ invariants ν$_0$;(ν$_1$ν$_2$ν$_3$) as proposed by Fu and Kane **[91]**. As found in **Refs. 9 and 15**, the products of the parities eigenvalues at all TRIM points of the primitive unit cell classify bulk α-GaGeTe as a strong Dirac TSM with ν$_0$;(ν$_1$ν$_2$ν$_3$)= 1;(111) that originates from a band inversion of the bulk valence and conduction bands at the Z point of the BZ.

On the other hand, the calculated band structure of β-GaGeTe shows a linear dispersion of both conduction and valence bands along the Γ-A direction as expected for a Dirac point. The calculations of the four topological $Z_2$ invariants for β-GaGeTe show that it is a weak topological semimetal with $v_0;(v_1v_2v_3)$= 0;(001) that shows a band inversion of the bulk valence and conduction bands at the Γ point of the BZ. Curiously, we have noted that the calculated band structure of β-GaGeTe shows a bandgap that is intermediate between those calculated for two compounds with the same space group as β-GaGeTe, non-centrosymmetric 3D hexagonal LiZnSb **[92]** and LiZnBi **[93]**. In fact, Dirac points have been predicted for LiZnBi **[93]** and both Dirac and Weyl points have been predicted for the intermediate LiZnSb$_x$Bi$_{1-x}$ Dirac semimetal **[94]**. Therefore, β-GaGeTe with point group $C_{6v}$ satisfies all the criteria to be a non-centrosymmetric TSM, as recently discussed **[94]**.

To finish, we must comment that TSMs without inversion symmetry have been much less studied than centrosymmetric TSMs. Theoretical predictions of material candidates for the former are comparatively rare **[94,95]** and experimental studies are even more scarce **[94-96]**. Therefore, the present discovery of layered non-centrosymmetric β-GaGeTe as a weak TSM is an important addition to the portfolio of TSMs and provides a framework for the study of quantum non-linear Hall effect, non-linear photocurrent, and non-linear optical properties in these compounds **[94-96]**.

**Conclusions**

We have reported a joint experimental and theoretical study of bulk topological semimetal GaGeTe by means of polarized Raman scattering measurements at room conditions under resonant and non-resonant excitation conditions combined with powder X-ray diffraction measurements and *ab initio* calculations. Unexpectedly, we have identified at least two different polytypes of layered GaGeTe in as-grown samples despite only one was supposed to exist. Apart from the centrosymmetric phase (here named α-GaGeTe) with s.g. *R-3m*, No. 166 and reported in the literature, we have found a non-centrosymmetric phase (here named β-GaGeTe) with s.g. *P6$_3$mc*, No. 186 that was not previously known. We have provided the structural parameters and vibrational frequencies at room conditions of both polytypes. Moreover, we have also shown that *ab initio* calculations also suggest the possibility of observing a third polytype of GaGeTe (centrosymmetric γ-GaGeTe) with s.g. *P-3m1* (No. 164) near room conditions; i.e. with the same symmetry of each of the monolayers of α-GaGeTe and β-GaGeTe but with a different internal atomic ordering. Correspondingly, we have also provided the theoretical structural parameters and vibrational frequencies at room conditions of γ-GaGeTe for its possible future experimental identification. These results show that GaGeTe is a much more complex material than previously thought and it opens the door to the control of the polytypism in GaGeTe, that can lead to engineer complex vdW heterostructures with the combination of only three elements (Ga, Ge, and Te).

We have also identified the experimental Raman-active modes of α-GaGeTe and β-GaGeTe and found that the two mid-frequency Raman-active modes of α-GaGeTe show a Fermi resonance. This resonance is observed in non-resonant Raman scattering measurements, but it is not observed in resonant Raman scattering measurements. To our knowledge, this on/off Fermi resonance effect has not been observed in solids before. On the other hand, several low-frequency and mid-frequency Raman-active modes of bulk β-GaGeTe are not observed in the RS spectrum likely due to weak Raman scattering cross-sections related to strong anharmonic scattering due to a large phonon density of states.

From lattice dynamics calculations, we have predicted an ultralow lattice thermal conductivity for both bulk α-GaGeTe and β-GaGeTe. Our results for α-GaGeTe are in good agreement with recent calculations in α-GaGeTe monolayers and in other topological materials. Therefore, both bulk α-GaGeTe and β-GaGeTe are potential candidates for highly efficient thermoelectric materials. Control of the deposition of the two polymorphs could lead to the engineering of heterostructures formed by alternate layers of both α and β polytypes that would allow thermoelectric modules of a single material with an ultralow lattice thermal conductivity that would be even smaller than those shown by individual polytypes.

Finally, we have calculated the electronic band structure of bulk α-GaGeTe and β-GaGeTe at room conditions and evaluated their topological properties. Both polymorphs are TSMs, being the gaps in bulk β-GaGeTe smaller than those in bulk α-GaGeTe. While centrosymmetric α-GaGeTe is a Dirac TSM, β-GaGeTe is a non-centrosymmetric weak TSM. Consequently, they are promising TSMs to find exotic fermions upon gap closure, which can be tuned using appropriate doping or even by applying moderate hydrostatic or uniaxial pressures. Moreover, it is known that breaking inversion symmetry in TSMs can lift the spin degeneracy through Rashba spin splitting and is important for many exotic phenomena, like topological ferroelectriciy, pyroelectricity, and multiferroicity. In particular, non-centrosymmetry is essential for the separation of Weyl points with opposite chiral charge. Therefore, GaGeTe offers the opportunity of combining centrosymmetric and non-centrosymmetric TSM properties in a single material.

In summary, we have clarified that there are at least two polytypes in commonly as-grown samples of GaGeTe. Therefore, a proper characterization of this material must be done when working with it to measure its intrinsic properties and use it in electronic and optoelectronic devices. TSM GaGeTe is a unique material to design and implement complex van der Waals heterostructures, with up to three different centrosymmetric and non-centrosymmetric phases, having the same chemical composition. These heterostructures can be implemented for fundamental studies and in a number of technological applications, including spintronics, thermoelectrics, solar cells, field-effect transistors, photodetectors, and devices for topological applications. We hope our work will help researchers to differentiate between the different polytypes of TSM GaGeTe and will stimulate further studies on these polytypes in bulk and 2D form.

**Acknowledgments**

This publication is part of the project MALTA Consolider Team network (RED2018-102612-T), financed by MINECO/AEI/10.13039/501100003329; by I+D+i projects PID2019-106383GB-41/42/43 financed by MCIN/AEI/10.13039/501100011033; and by project PROMETEO/2018/123 (EFIMAT) financed by Generalitat Valenciana. E.B. would like to thank the Universitat Politècnica de València for his postdoctoral contract (Ref. PAID-10-21). AHR was supported by the U.S. Department of Energy (DOE), Office of Science, Basic Energy Sciences under award DE-SC0021375. We also acknowledge the computational resources awarded by XSEDE, a project supported by National Science Foundation grant number ACI-1053575. The authors also acknowledge the support from the Texas Advances Computer Center (with the Stampede2 and Bridges supercomputers). E.L.d.S would like to acknowledge the Network of Extreme Conditions Laboratories (NECL), financed by FCT and co-financed by NORTE 2020, through the program Portugal 2020 and FEDER; the High Performance Computing Chair - a R&D infrastructure (based at the University of Évora; PI: M. Avillez); and for the computational support provided by the HPC center OBLIVION-U.Évora to perform the lattice thermal conductivity calculations. A.L. and D.E. would like to thank the Generalitat Valenciana for the Ph.D. Fellowship no. GRISOLIAP/2019/025.

**Tables**

**Table 1.** Experimental (ex.) and theoretical (th.) Raman-active (gerade, g) and infrared-active (ungerade, u) frequencies, ω, at ambient conditions for α-GaGeTe. The relative differences between experimental and theoretical frequencies, $R_\omega$, are shown. Experimental data from Ref. 14 are also included for comparison.

| Mode | ω (th.)[a] (cm$^{-1}$) | ω (th.)[b] (cm$^{-1}$) | ω (ex.) (cm$^{-1}$) | $R_\omega$[a] (%) | $R_\omega$[b] (%) | ω[c] (cm$^{-1}$) | $R_\omega$[c] (%) |
|---|---|---|---|---|---|---|---|
| $E_g^1$ | 40.1 | 39.4 | 41(1) | 2.2 | 3.9 | - | - |
| $E_u^1$ | 52.5 | 52.7 | | | | | |
| $A_{1g}^1$ | 76.1 | 76.9 | 78(1) | 2.4 | 1.4 | 64 | 19 |
| $A_{2u}^1$ | 138.6 | 136.7 | | | | | |
| $E_g^2$ | 177.0 | 172.6 | 177(2) | 0.0 | 2.5 | 177 | 0 |
| $E_u^2$ | 177.1 | 173.1 | | | | | |
| $A_{1g}^2$ | 197.9 | 197.1 | 205(2) | 3.5 | 3.9 | 166 | 19 |
| $A_{2u}^2$ | 267.3 | 263.5 | 266(1) | 0.5 | 0.9 | 263 | 2 |
| $E_g^3$ | 276.0 | 268.1 | 279(3) | 1.1 | 3.9 | 284 | 2 |
| $A_{1g}^3$ | 282.6 | 282.1 | 296(1) | 4.5 | 4.7 | 295 | 4 |

[a] PBEsol, [b] PBE-D3, [c] Ref. 14

**Table 2.** Experimental (theoretical PBEsol) atomic coordinates of the *R-3m* (s.g. No. 166) structure of α-GaGeTe as obtained from Rietveld refinement. The experimental (theoretical PBEsol) [theoretical PBE-D3] lattice parameters of the hexagonal unit cell are a = 4.0480 Å (4.0495 Å) [4.1026 Å], b = 4.0480 Å (4.0495 Å) [4.1026 Å], c = 34.7340 Å (34.4336 Å) [34.7826 Å], and $V_0$ = 492.8800 Å$^3$ (489.0000 Å$^3$) [506.9999 Å$^3$]. Experimental Rietveld R-factors are: $R_p$ = 7.27, $R_{wp}$ = 5.24, $R_{exp}$ = 9.81, Chi$^2$ = 1.873.

| Atom | Wyckoff site | x | y | z |
|---|---|---|---|---|
| Ga | 6c | 0 | 0 | 0.2550 (0.2513) |
| Ge | 6c | 0 | 0 | 0.3224 (0.3218) |
| Te | 6c | 0 | 0 | 0.1173 (0.1188) |

**Table 3.** Experimental (ex.) and theoretical (th.) Raman- and infrared-active frequencies, ω, at ambient conditions for β-GaGeTe. The relative differences between experimental and theoretical frequencies, $R_\omega$, are shown. Experimental data from Ref. 14 are also included for comparison.

| Mode | ω (th.)[a] (cm$^{-1}$) | ω (th.)[b] (cm$^{-1}$) | ω (ex.) (cm$^{-1}$) | ω (ex.)[c] (cm$^{-1}$) | $R_\omega$[a] (%) | $R_\omega$[b] (%) |
|---|---|---|---|---|---|---|
| $E_2^1$ | 16.2 | 17.2 | | | | |
| $E_2^2$ | 35.6 | 35.8 | | | | |
| $E_1^1$ | 39.1 | 39.2 | 35(1) | - | 11.7 | 11.8 |
| $E_1^2$ | 52.5 | 53.3 | 59(1) | | 11.0 | 9.7 |
| $E_2^3$ | 53.1 | 53.9 | | | | |
| $A_1^1$ | 75.7 | 77.2 | 66(1) | 64 | 14.7 | 17.0 |
| $A_1^2$ | 139.4 | 138.0 | 168(2) | 166 | 17.0 | 17.8 |
| $E_1^3$ | 176.3 | 173.0 | 177(3) | 177 | 0.2 | 2.1 |
| $E_2^4$ | 176.6 | 173.2 | 177(3) | | 0.2 | 2.1 |
| $E_1^4$ | 178.4 | 174.9 | 182(4) | | 1.9 | 3.9 |
| $E_2^5$ | 178.6 | 175.1 | 182(4) | | 1.9 | 3.9 |
| $A_1^3$ | 201.0 | 199.2 | 200(4) | | 0.5 | 0.4 |
| $A_1^4$ | 267.9 | 265.0 | 263(1) | 263 | 1.8 | 0.8 |
| $E_1^5$ | 274.6 | 267.9 | 285(1) | 284 | 3.6 | 6.0 |
| $E_2^6$ | 274.7 | 268.0 | 285(1) | | 3.6 | 6.0 |
| $A_1^5$ | 285.6 | 283.8 | 295(1) | 295 | 3.2 | 3.8 |

[a] PBEsol, [b] PBE-D3, [c] Ref. 14

**Table 4.** Experimental (theoretical PBEsol) atomic coordinates of the $P6_3mc$ (s.g. No. 186) structure of β-GaGeTe. The experimental (theoretical PBEsol) [theoretical PBE-D3] lattice parameters of the hexagonal unit cell are a = 4.0379 Å (4.0451 Å) [4.0940 Å], b = 4.0379 Å (4.0451 Å) [4.0940 Å], c = 22.1856 Å (23.0047 Å) [23.1474 Å], and $V_0$ = 313.2658 Å$^3$ (326.0046 Å$^3$) [335.9999 Å$^3$]. Conventional Rietveld R-factors are: $R_p$ = 9.72, $R_{wp}$ = 7.84, $R_{exp}$ = 3.76, Chi$^2$ = 2.087.

| atom | Wyckoff site | x | y | z |
|---|---|---|---|---|
| Ga1 | 2a | 0 | 0 | 0.1270 (0.1271) |
| Ga2 | 2b | 1/3 | 2/3 | 0.3729 (0.3729) |
| Ge1 | 2a | 0 | 0 | 0.2327 (0.2327) |
| Ge2 | 2b | 1/3 | 2/3 | 0.2672 (0.2673) |
| Te1 | 2b | 1/3 | 2/3 | 0.5727 (0.5720) |
| Te2 | 2b | 1/3 | 2/3 | 0.9274 (0.9281) |

**Figure captions**

**Figure 1.** Crystal structure of bulk α-GaGeTe (left), β-GaGeTe (center), and γ-GaGeTe (right).

**Figure 2.** Unpolarized and polarized (parallel and cross polarizations) RS spectra of bulk α-GaGeTe (a) and bulk β-GaGeTe (b) at room conditions excited under non-resonant (632.8 nm) and resonant (532 nm) conditions. The bottom marks correspond to the theoretically predicted frequencies for both polytypes (black and blue for E and A modes in α-GaGeTe, respectively, and black, red, and blue for $E_1$, $E_2$, and $A_1$ modes in β-GaGeTe, respectively).

**Figure 3.** Experimental (black line) XRD patterns of as-grown GaGeTe sample with (a) the *R-3m* structure of α-GaGeTe and (b) the *R-3m* and *P6$_3$mc* structures of α-GaGeTe and β-GaGeTe, respectively. The Rietveld refinement fit (red lines) as well as the residuals of the refinements (blue lines) are also shown.

**Figure 4.** Theoretical total energy vs volume for the different polymorphs of GaGeTe. α-GaGeTe (s.g. *R-3m*), β-GaGeTe (s.g. *P6$_3$mc*), γ-GaGeTe (s.g. *P-3m1*), and GaGeTe-mono (s.g. *P-3m1*).

**Figure 5.** Detail of the effect of pressure on the RS spectra of α-GaGeTe in the mid-frequency region (of the $E_g^2$ and $A_{1g}^2$ modes) excited with the red laser (a) and the green laser (b). Idem on the RS spectra of β-GaGeTe in the mid-frequency region (of the $A_1^2$, $E_1^3$, $E_1^4$, and $A_1^3$ modes) excited with the red laser (d) and the green laser (e). Experimental (symbols) and theoretical (lines) pressure dependence of the mid-frequency Raman-active modes in bulk α-GaGeTe (c) and bulk β-GaGeTe (f). Circles and squares correspond to experimental data obtained withe the red and green lasers, respectively. Filled symbols in (c) correspond to experimental data of the main Raman-active modes measured with the green laser under resonant excitation at low pressures and under weak Fermi resonance at high pressures. Open symbols in (c) correspond to experimental data of weak relative maxima of the broad bands under Fermi resonance (see red marks in (a) and (b)).

**Figure 6.** Lattice thermal conductivity ($\kappa_L$) for bulk α-GaGeTe (top) and bulk β-GaGeTe (bottom) as a function of temperature along the main directions of the hexagonal unit cell: layer plane (xx = yy) and c axis (zz).

**Figure 7.** Top: Theoretical electronic band structure of bulk α-GaGeTe along the main points (U-Γ-Z-F-Γ-L) of the BZ. Bottom: Theoretical electronic band structure of bulk β-GaGeTe along the main points (Γ-M-K-Γ-A-L-H-A) of the BZ.

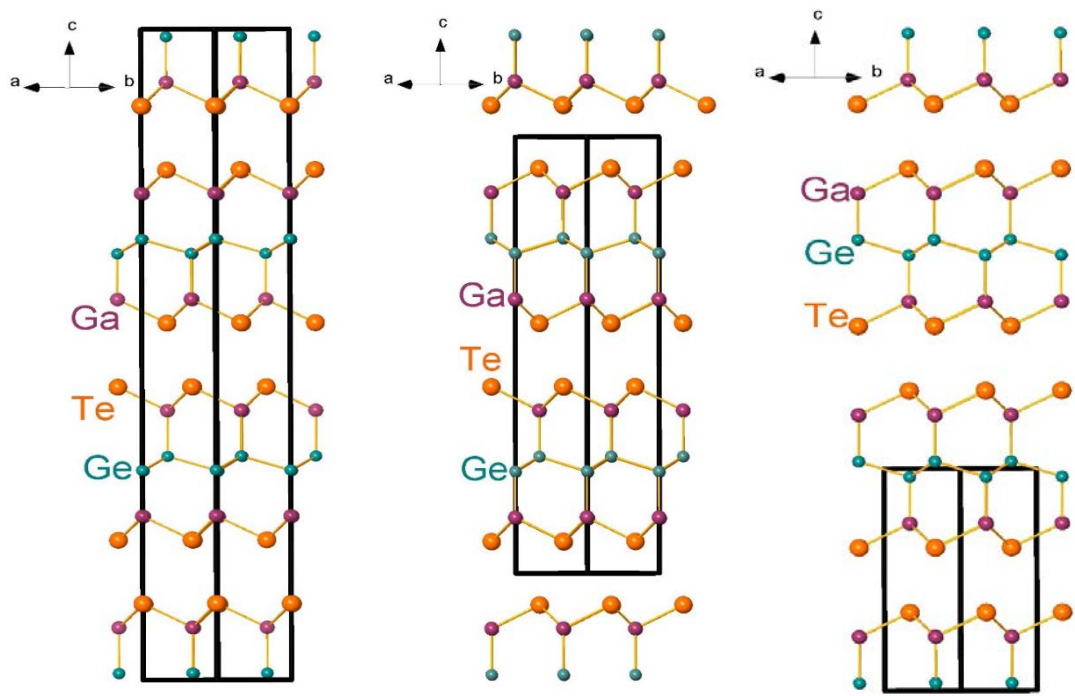

Figure 1

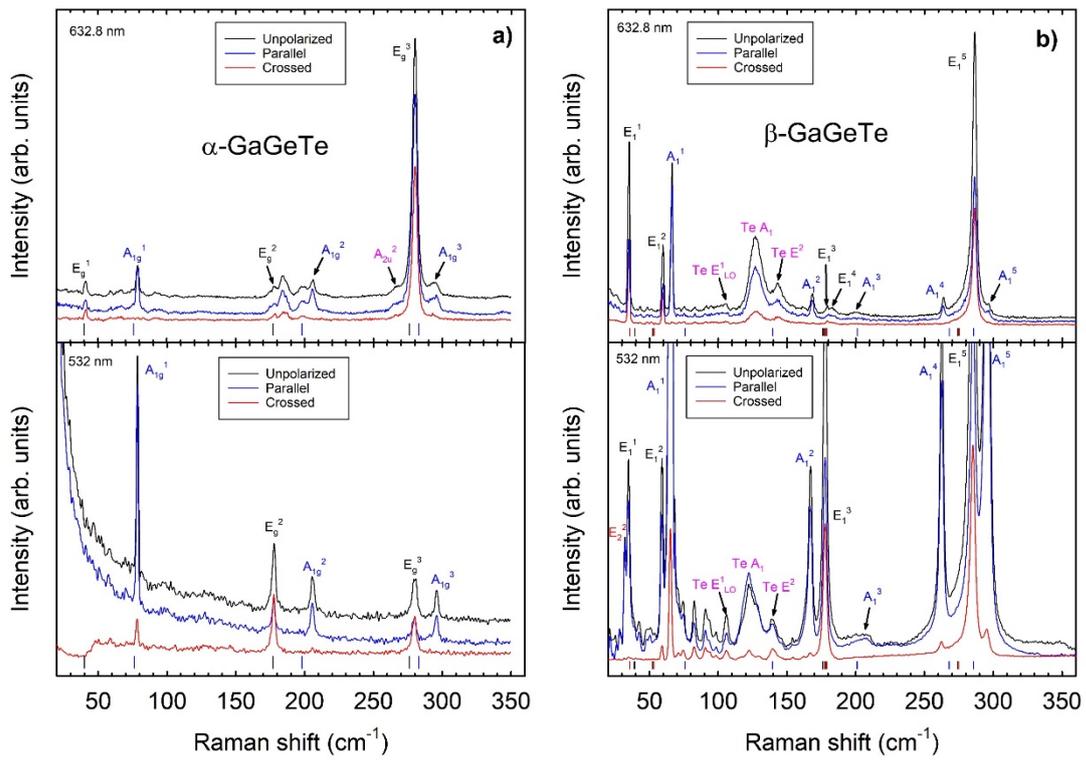

Figure 2

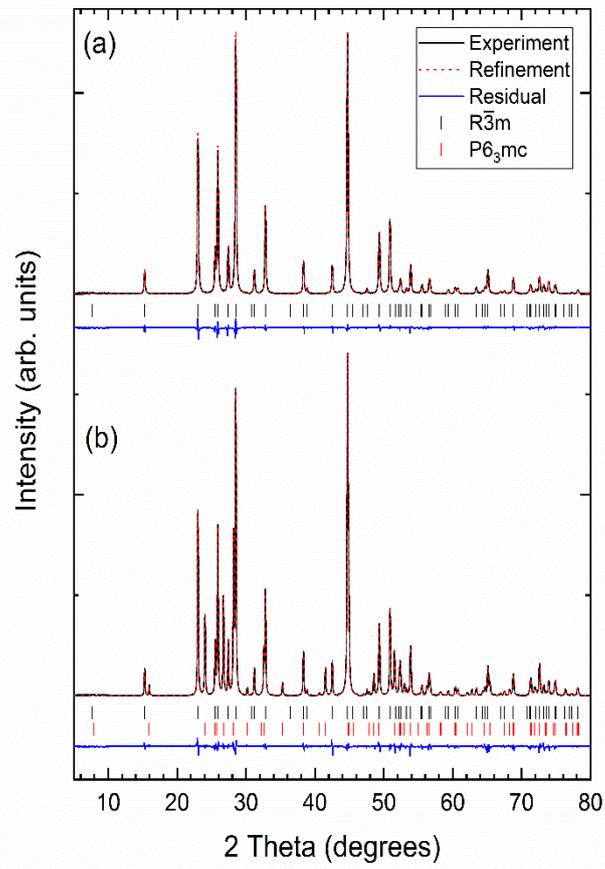

Figure 3

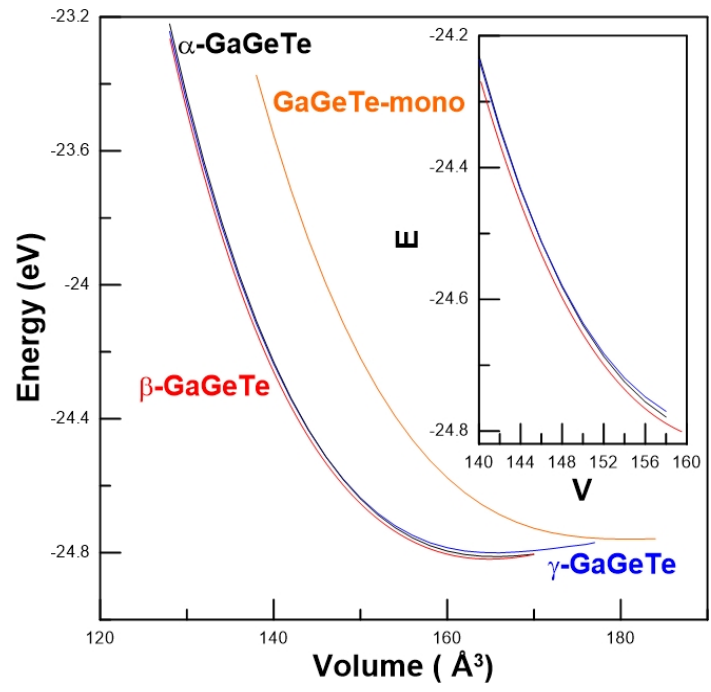

Figure 4

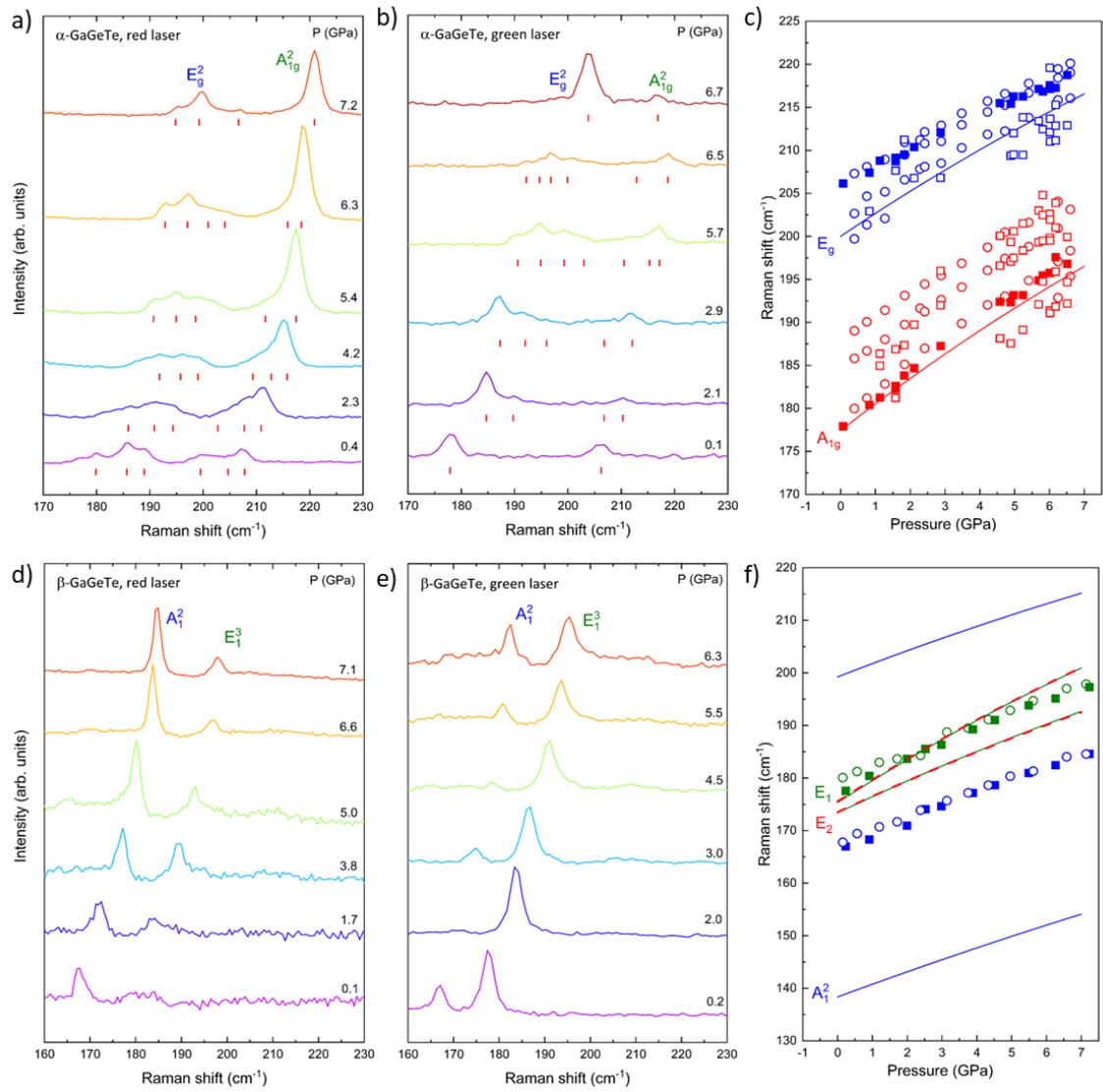

Figure 5

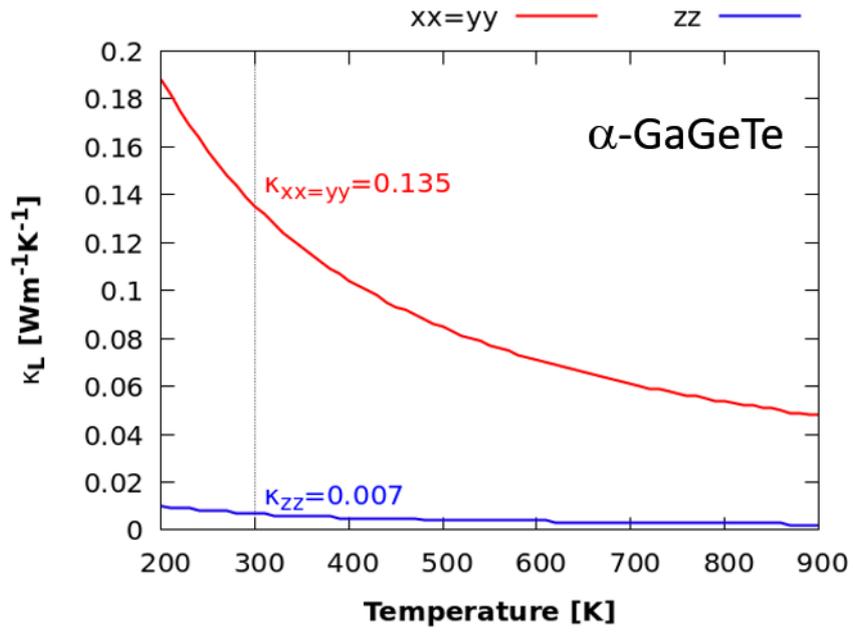

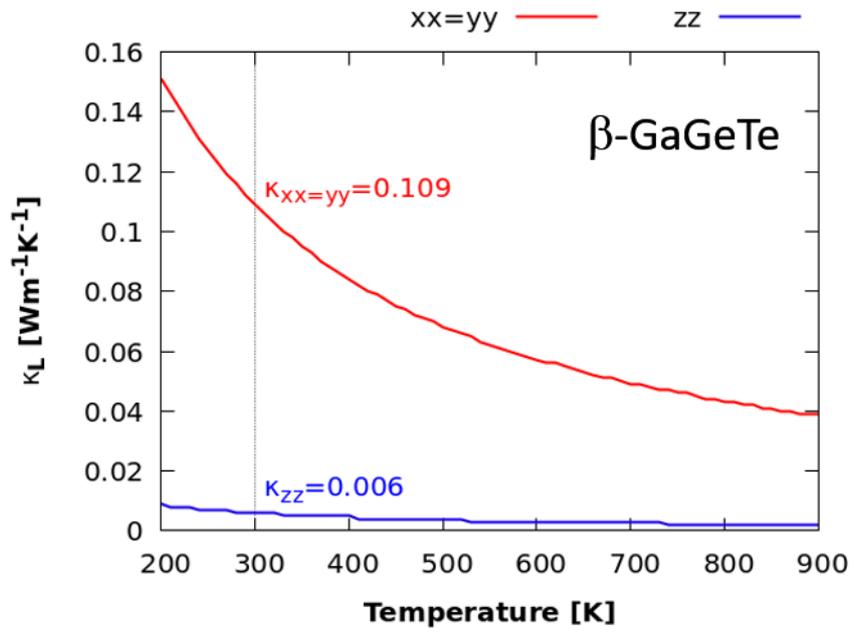

Figure 6

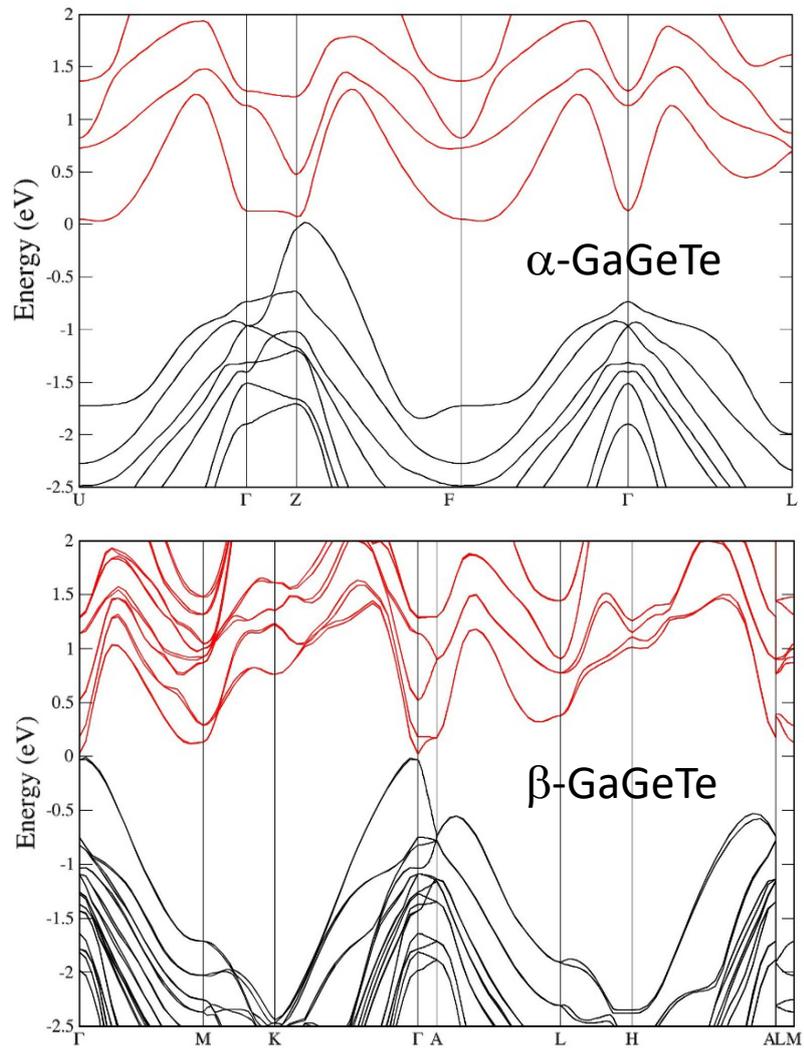

Figure 7

# Supplementary Material of

# Layered topological semimetal GaGeTe: New polytype with non-centrosymmetric structure

S. Gallego-Parra,[1] E. Bandiello,[1] A. Liang,[2] E. Lora da Silva,[3] P. Rodríguez-Hernández,[4] A. Muñoz,[4] S. Radescu,[4] A.H. Romero,[5] C. Drasar,[6] D. Errandonea,[2] and F. J. Manjón[1,*]

[1] Instituto de Diseño para la Fabricación y Producción Automatizada, MALTA Consolider Team, Universitat Politècnica de València, 46022 Valencia, Spain

[2] Departamento de Física Aplicada-ICMUV, MALTA Consolider Team, Universitat de València, 46100 Burjassot, Spain

[3] IFIMUP, Departamento de Física e Astronomia, Faculdade de Ciências, Universidade do Porto, 4169-007 Porto, Portugal

[4] Departamento de Física, Instituto de Materiales y Nanotecnología, MALTA Consolider Team, Universidad de La Laguna, La Laguna, 38205 Tenerife, Spain

[5] Department of Physics and Astronomy, West Virginia University, Morgantown, West Virginia 26506-6315, USA

[6] Faculty of Chemical Technology, University of Pardubice, Pardubice 532 10, Czech Republic

* corresponding author: fjmanjon@fis.upv.es

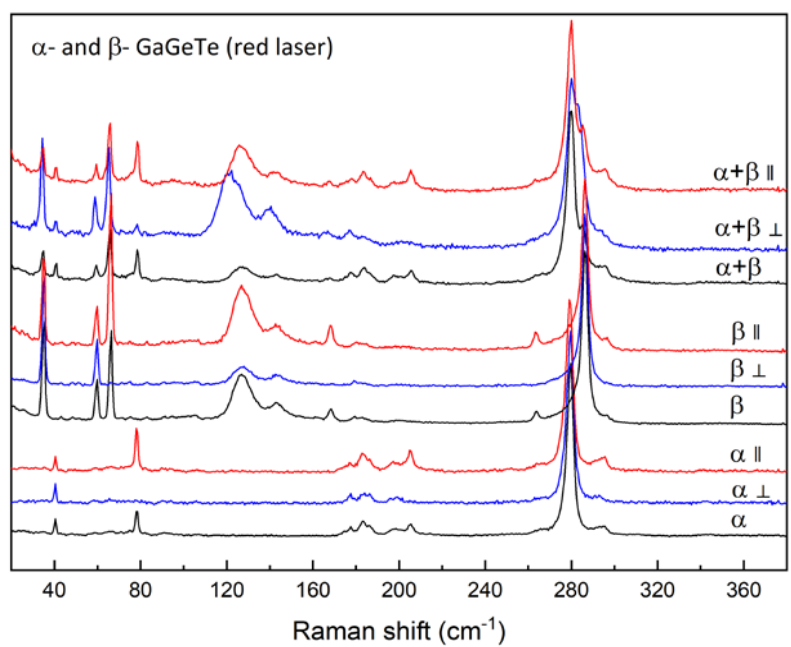

**Figure S1.** Unpolarized and polarized (parallel and cross polarizations) RS spectra of α-GaGeTe β-GaGeTe and a mixture of both polytypes at room conditions excited under non-resonant (632.8 nm) conditions.

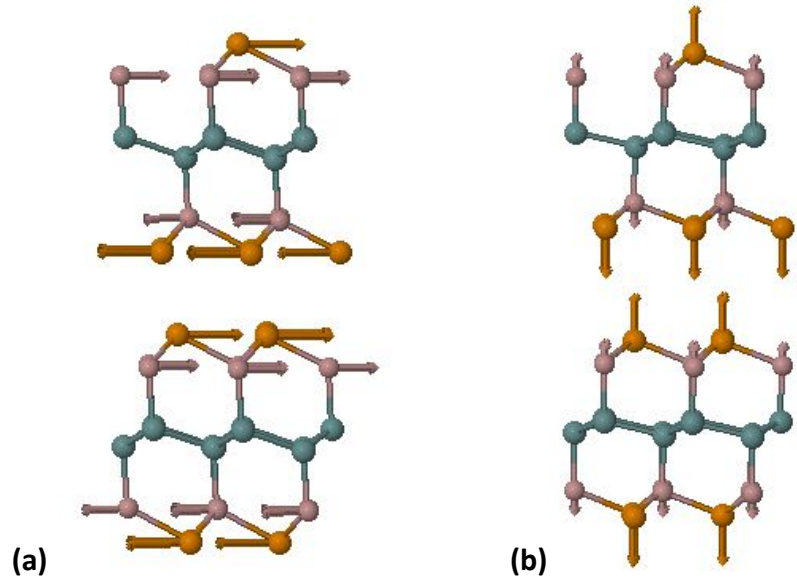

**Figure S2.** Atomic motion of $E_g^1$ (a) and $A_{1g}^1$ (b) vibrational modes of α-GaGeTe predicted at 40.1 and 76.1 cm$^{-1}$. Ga, Ge, and Te atoms are depicted in pink, gray, and orange colors. These two modes show Ge atoms at rest and Ga and Te atoms in motion. These two modes are the rigid layer modes of α-GaGeTe where each layer vibrate against the neighboring layer. $E_g^1$ is the transversal or shear layer mode (with a pure Ga-Ge bending mode contribution) and $A_{1g}^1$ is the longitudinal or compressional layer mode (with small contribution of a Ga-Ge symmetric stretching mode).

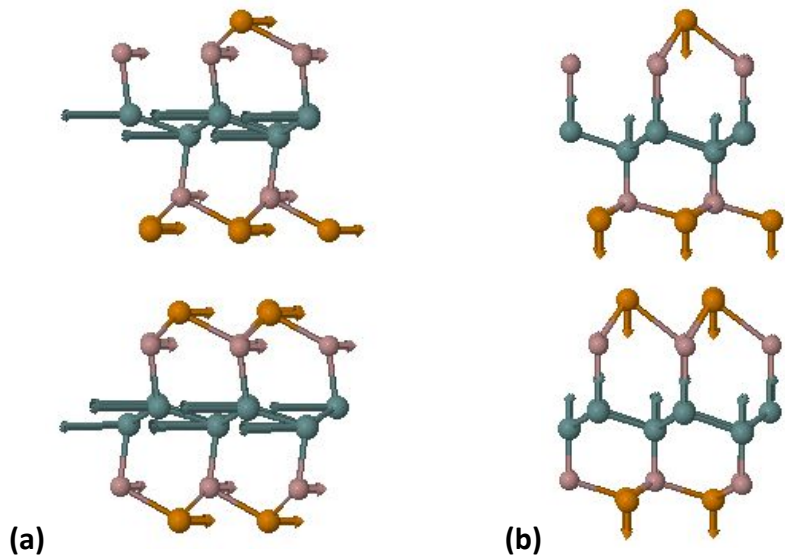

**Figure S3.** Atomic motion of $E_u^1$ (a) and $A_{2u}^1$ (b) vibrational modes of α-GaGeTe predicted at 52.5 and 138.6 cm$^{-1}$. Ga, Ge, and Te atoms are depicted in pink, gray, and orange colors. These two modes show Ge atoms vibrating out-of-phase with respect to Ga and Te atoms. These two modes are rigid intra-layer modes of α-GaGeTe, where the germanene sublayer vibrate against the other two Ga-Te sublayers. $E_u^1$ is the transversal or shear intra-layer mode (a pure Ga-Ge bending mode) and $A_{2u}^1$ is the longitudinal or compressional intra-layer mode (almost a pure Ga-Te bending mode with small contribution of a Ga-Ge asymmetric stretching mode).

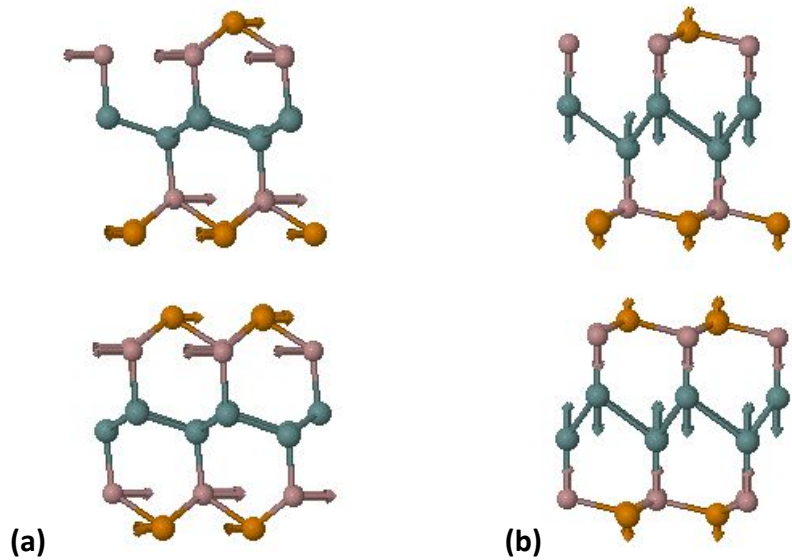

**Figure S4.** Atomic motion of $E_g^2$ (a) and $A_{1g}^2$ (b) vibrational modes of α-GaGeTe predicted at 177.0 and 197.9 cm$^{-1}$. Ga, Ge, and Te atoms are depicted in pink, gray, and orange colors. The first mode shows Ge atoms at rest and Ga and Te atoms in motion. The $E_g^2$ mode is a mixture of a Ga-Ge bending mode, a Ga-Te stretching mode, and a weak interlayer shear mode. The $A_{1g}^2$ mode is a mixture of a Ge-Ge bending mode, a Ga-Te bending mode, and a very weak interlayer compressional mode due to the small amplitude of Te vibrations along the *c*-axis in comparison with the $A_{1g}^1$ mode.

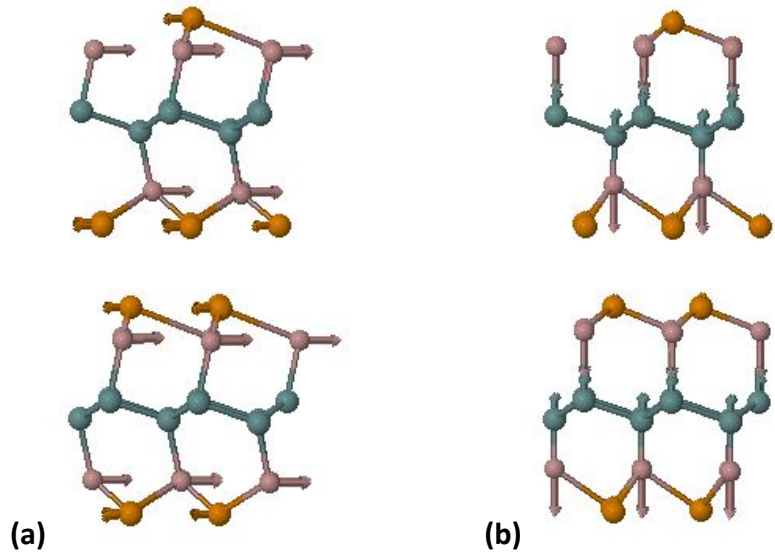

**Figure S5.** Atomic motion of $E_u^2$ (a) and $A_{2u}^2$ (b) vibrational modes of α-GaGeTe predicted at 177.1 and 267.3 cm$^{-1}$. Ga, Ge, and Te atoms are depicted in pink, gray, and orange colors. The first mode shows Ge atoms at rest and Ga and Te atoms in motion. The $E_u^2$ mode is a mixture of a Ga-Ge bending mode and a Ga-Te stretching mode. The $A_{2u}^2$ mode is a mixture of a Ga-Ge asymmetric stretching mode and a Ga-Te bending mode.

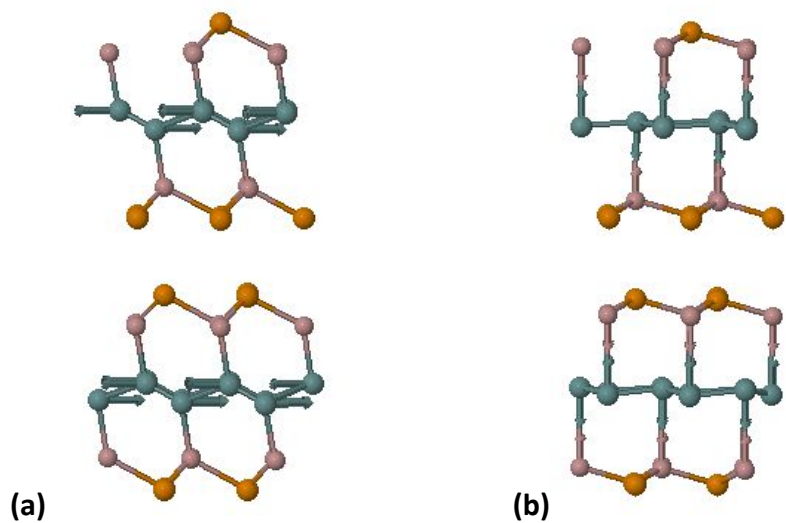

**Figure S6.** Atomic motion of $E_g^3$ (a) and $A_{1g}^3$ (b) vibrational modes of α-GaGeTe predicted at 276.0 and 282.6 cm$^{-1}$. Ga, Ge, and Te atoms are depicted in pink, gray, and orange colors. These two modes show Te atoms at rest and Ga and Ge atoms in motion. In the $E_g^3$ mode, neighbor Ge atoms vibrate out-of-phase. This mode is a mixture of a Ge-Ge stretching mode and a Ga-Ga bending mode. On the other hand, the $A_{1g}^3$ mode is a mixture of a Ga-Ge symmetric stretching mode and a Ge-Ge bending mode. The close mass of Ga and Ge leads to a similar frequency of both stretching modes.

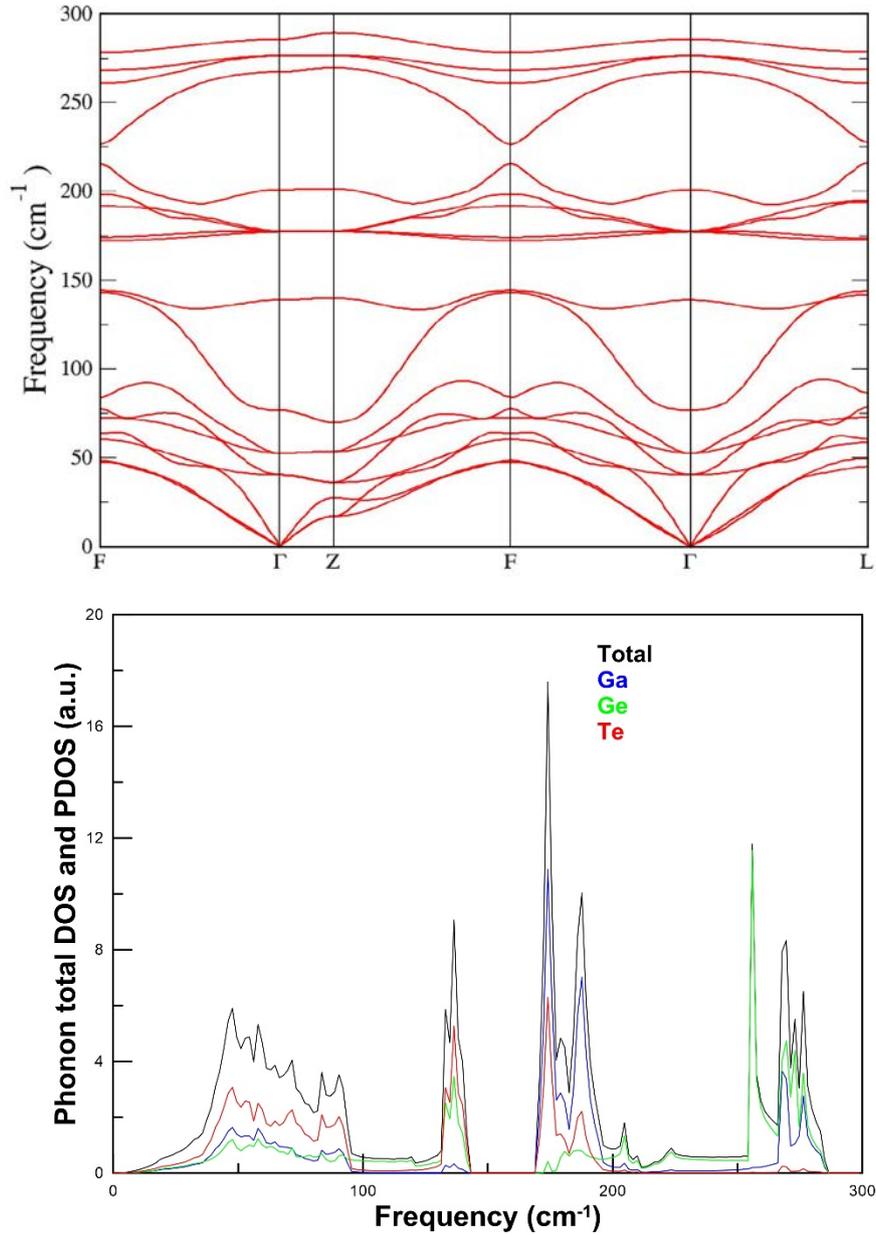

**Figure S7.** Top: Phonon dispersion curves of α-GaGeTe. Bottom: Total and partial (atom-projected) one-phonon density of states of α-GaGeTe.

The phonon dispersion curves of α-GaGeTe shows the typical phonon branches of a tetrahedrally-coordinated compound, like ZnO **[S1,S2]**, with phonon branches: i) related to transversal acoustic (TA) phonons (up to 75 cm$^{-1}$); ii) related to longitudinal acoustic (LA) phonons (from 75 up to 145 cm$^{-1}$); and iii) related to transversal optic (TO) and longitudinal optic (LO) phonons (from 165 up to 290 cm$^{-1}$). This means that TA phonons are in the region of the $E_g^1$ and $E_u^1$ modes, LA phonons include the region of $A_{1g}^1$ and $A_{2u}^1$ modes, and TO and LO phonons are mixed in the mid-frequency region of $E_g^2$, $E_u^2$ and $A_{1g}^2$ modes and in the high-frequency region of $A_{2u}^2$, $E_g^3$, and $A_{1g}^3$ modes. Additionally, we must note that there is a phonon gap between acoustic and optical branches (from 145 to 165 cm$^{-1}$). The most interesting feature of the phonon dispersion curves is the low-frequency values found for the TA and some LA modes at the Z point of the Brillouin zone. These low-frequency modes below 75 cm$^{-1}$ suggest a very low thermal conductivity in bulk GaGeTe along the *c*-axis, in agreement with what has been recently predicted for monolayer GaGeTe **[S3]**.

The one-phonon density of states shows that Te vibrations contribute to vibrational modes up to 200 cm$^{-1}$, while Ga and Ge contribute to all vibrational modes, with Ge modes contributing mostly to vibrations close to 135 cm$^{-1}$ and above 250 cm$^{-1}$ and Ga modes contributing mostly to vibrations between 165 and 200 cm$^{-1}$ and above 265 cm$^{-1}$. In the one-phonon DOS, the highest contribution comes from the mid-frequency modes $E_g^2$ and $E_u^2$ between 170 and 200 cm$^{-1}$ due to the low dispersion of the curves of these two modes along the BZ.

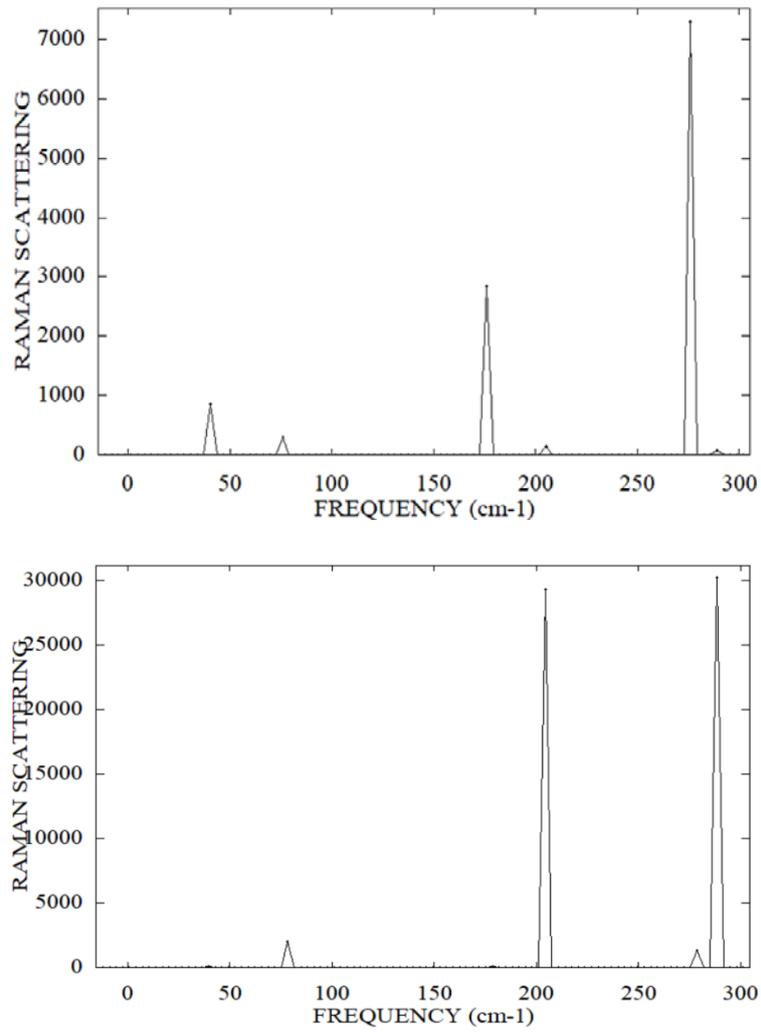

**Figure S8.** Simulated unpolarized RS spectrum of α-GaGeTe (top) and γ-GaGeTe (bottom).

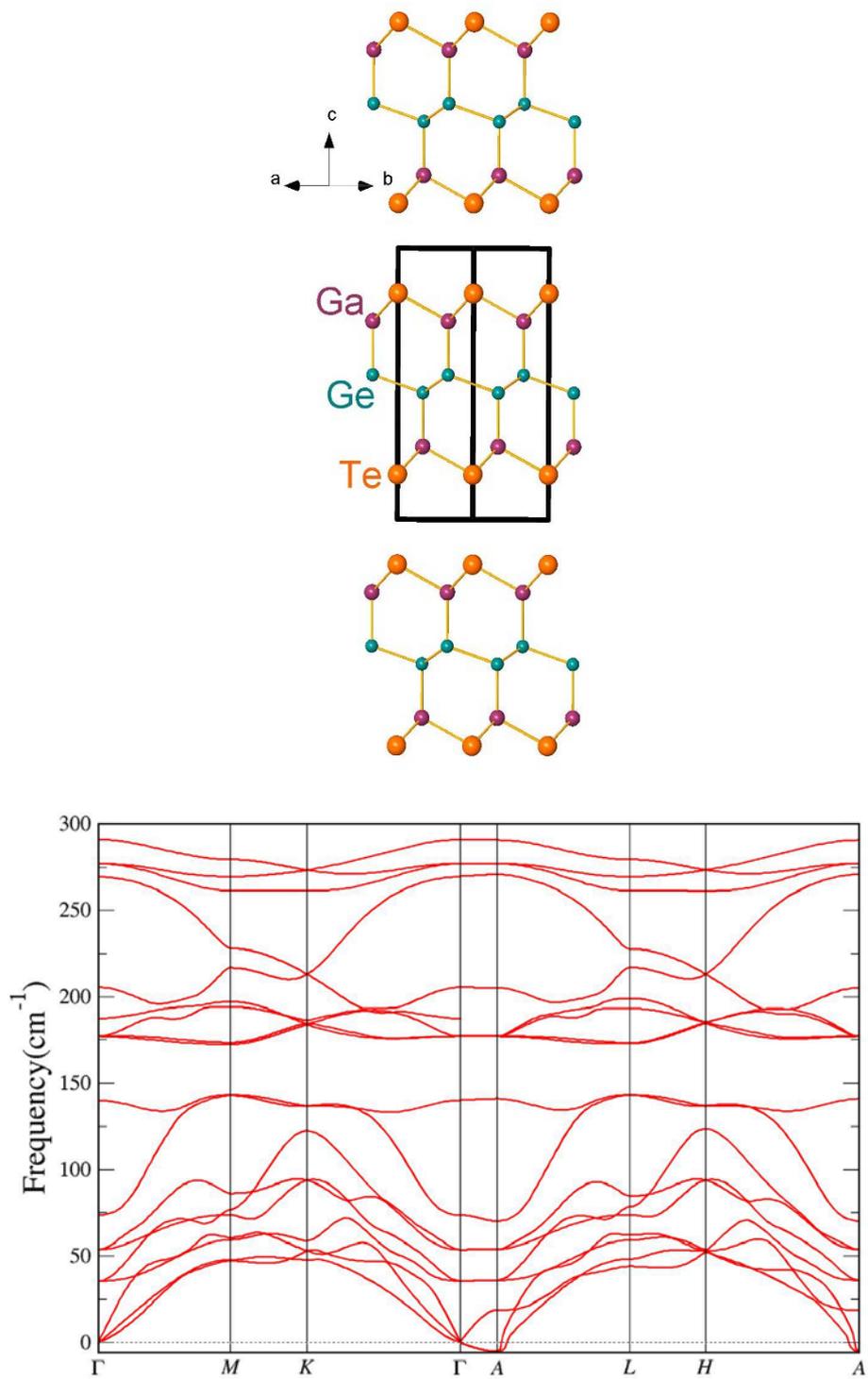

**Figure S9.** Detail of the crystalline structure of GaGeTe-mono (top) and theoretical phonon dispersion curves of GaGeTe-mono (bottom) along the main points (Γ—M—K—Γ—A—L—H—A) of the Brillouin zone. Note the imaginary frequencies near the A point.

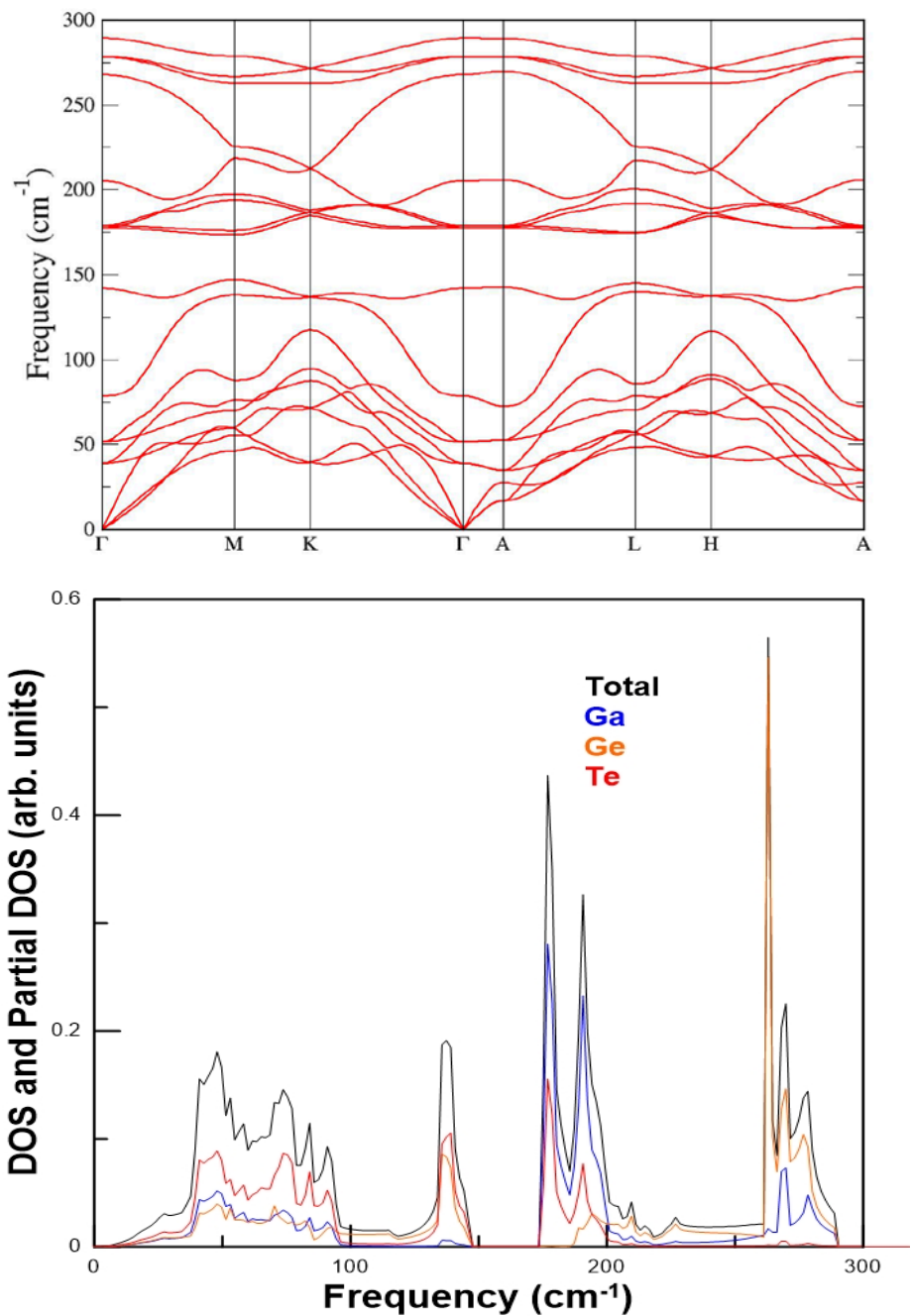

**Figure S10.** Top: Theoretical phonon dispersion curves of γ-GaGeTe along the main points (Γ—M—K—Γ—A—L—H—A) of the Brillouin zone. Bottom: Theoretical total and partial (atom-projected) one-phonon density of states of γ-GaGeTe.

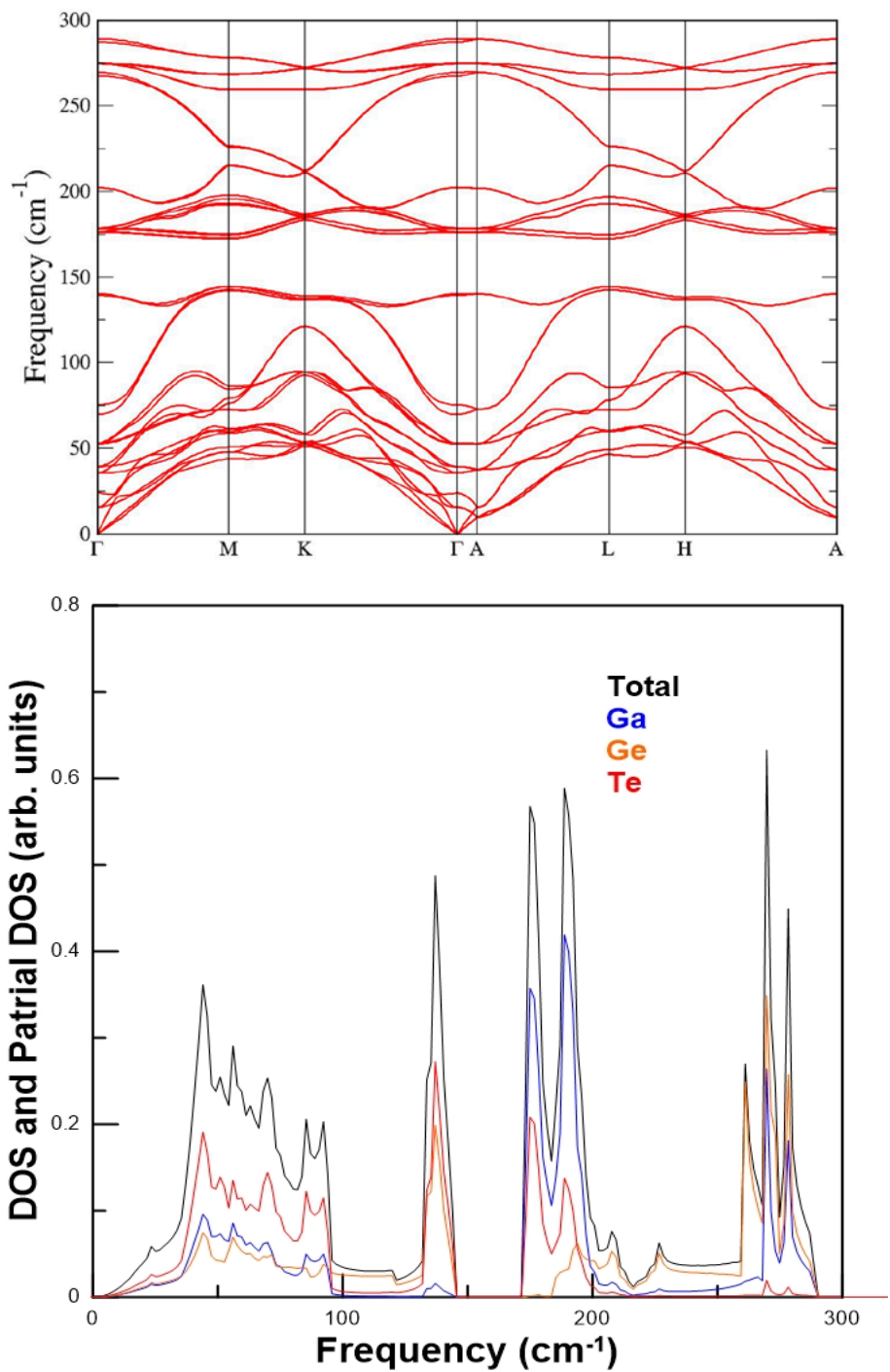

**Figure S11.** Top: Theoretical phonon dispersion curves along the main points (Γ—M—K—Γ—A—L—H—A) of the Brillouin zone of β-GaGeTe. Bottom: Theoretical total and partial (atom-projected) one-phonon density of states of β-GaGeTe.

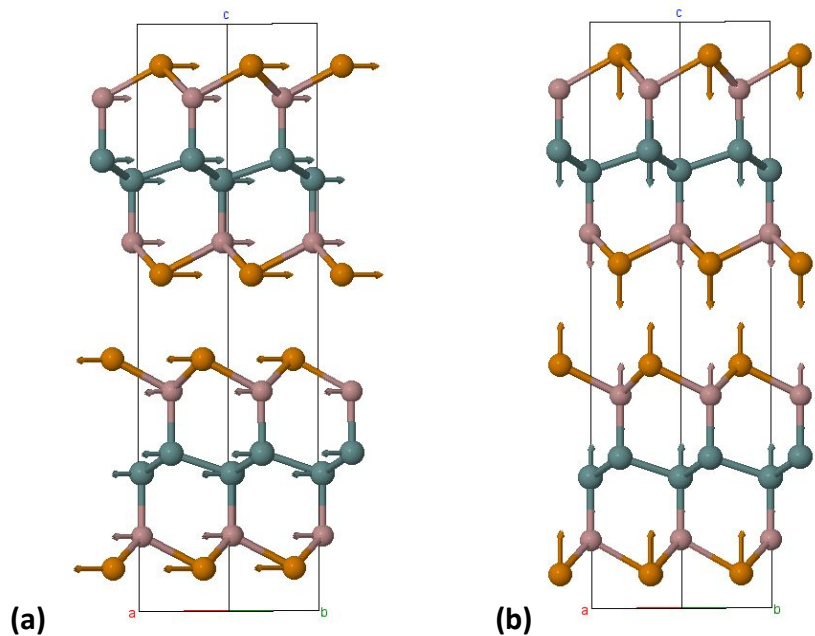

**Figure S12.** Atomic motion of $E_2^1$ (a) and $B_1^1$ (b) vibrational modes of β-GaGeTe predicted at 16.2 and 24.0 cm$^{-1}$. Ga, Ge, and Te atoms are depicted in pink, gray, and orange colors. These two modes are the rigid layer modes of β-GaGeTe where each layer vibrate against the neighboring layer. $E_2^1$ is the transversal or shear layer mode and $B_1^1$ is the longitudinal or compressional layer mode.

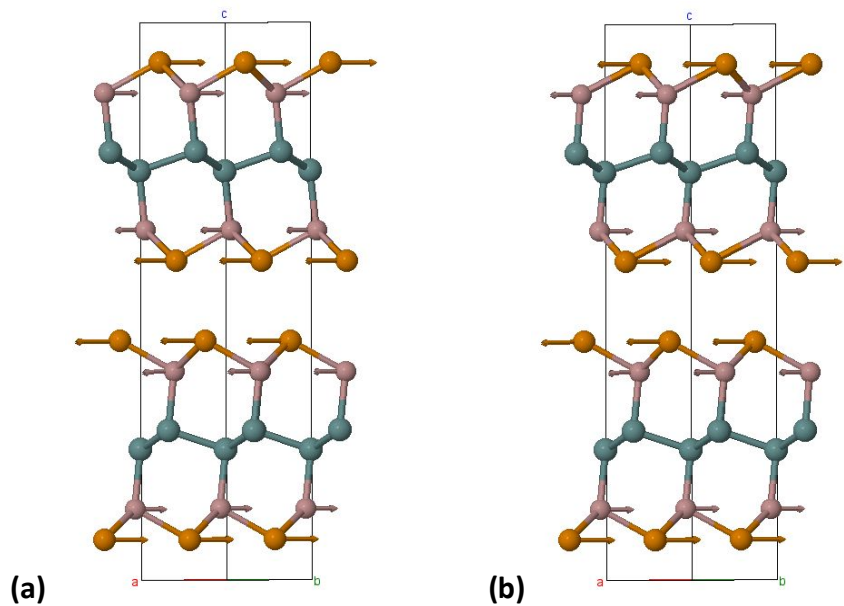

**Figure S13.** Atomic motion of $E_2^2$ (a) and $E_1^1$ (b) vibrational modes of β-GaGeTe predicted at 35.6 and 39.1 cm$^{-1}$. Ga, Ge, and Te atoms are depicted in pink, gray, and orange colors. These two modes are shear intra-layer modes of β-GaGeTe (pure Ga-Ge bending modes), where Ge atoms are at rest and the Ga-Te sublayers vibrate in opposite directions. In the $E_2^2$ mode, adjacent Ga-Te atoms of neighbour layers vibrate in-phase while in the $E_1^1$ mode, adjacent Ga-Te atoms of neighbour layers vibrate out-of-phase.

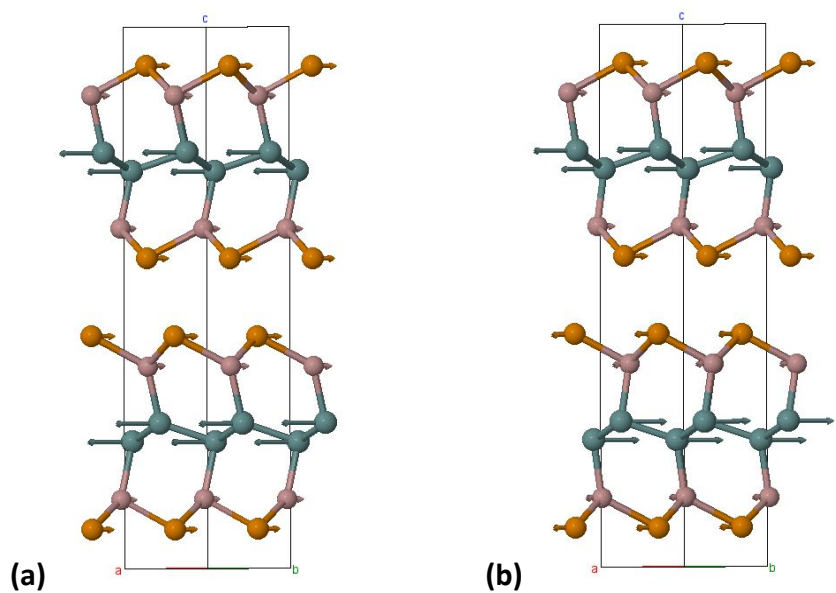

**Figure S14.** Atomic motion of $E_1^2$ (a) and $E_2^3$ (b) vibrational modes of β-GaGeTe predicted at 52.5 and 53.1 cm$^{-1}$. Ga, Ge, and Te atoms are depicted in pink, gray, and orange colors. These two modes are shear intra-layer modes of β-GaGeTe (also pure Ga-Ge bending modes), where Ge atoms vibrate in the opposite direction to the top and bottom Ga-Te sublayers. In the $E_1^2$ mode, adjacent Ga-Te atoms of neighbour layers vibrate in-phase while in the $E_2^3$ mode, adjacent Ga-Te atoms of neighbour layers vibrate out-of-phase.

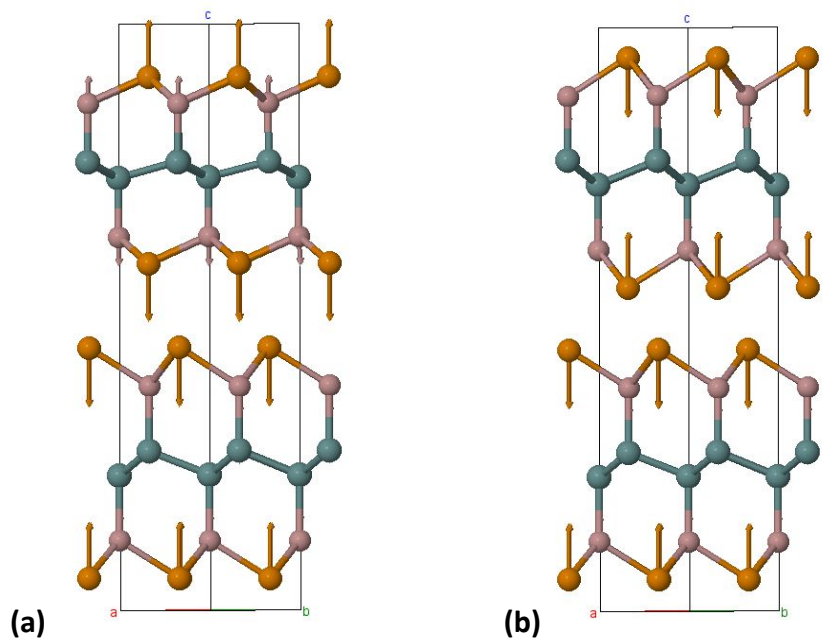

**Figure S15.** Atomic motion of $B_1^2$ (a) and $A_1^1$ (b) vibrational modes of β-GaGeTe predicted at 69.7 and 75.7 cm$^{-1}$. Ga, Ge, and Te atoms are depicted in pink, gray, and orange colors. These two modes are longitudinal intra-layer modes of β-GaGeTe where Ge atoms at rest and Ga and Te atoms vibrating against the Ge sublayer (partial Ga-Ge symmetric stretching mode). In the $B_1^2$ mode, the Ga-Te sublayers of neighbour layers vibrate in-phase, while in the $A_1^1$ mode the Ga-Te sublayers of neighbour layers vibrate out-of-phase.

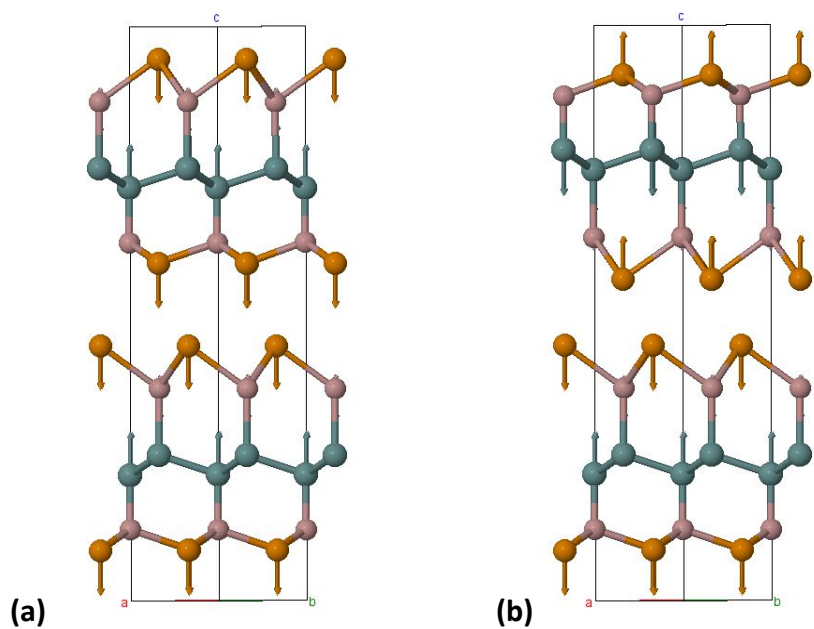

**Figure S16.** Atomic motion of $A_1^2$ (a) and $B_1^3$ (b) vibrational modes of β-GaGeTe predicted at 139.4 and 140.4 cm$^{-1}$. Ga, Ge, and Te atoms are depicted in pink, gray, and orange colors. These two modes are also longitudinal intra-layer modes of β-GaGeTe where Ge atoms vibrate in-phase with one Ga-Te sublayer and out-of-phase with the other Ga-Te sublayer (full Ga-Ge symmetric stretching mode). In the $A_1^2$ mode the Ga-Te sublayers of neighbour layers vibrate in-phase, while in the $B_1^3$ mode the Ga-Te sublayers of neighbour layers vibrate out-of-phase.

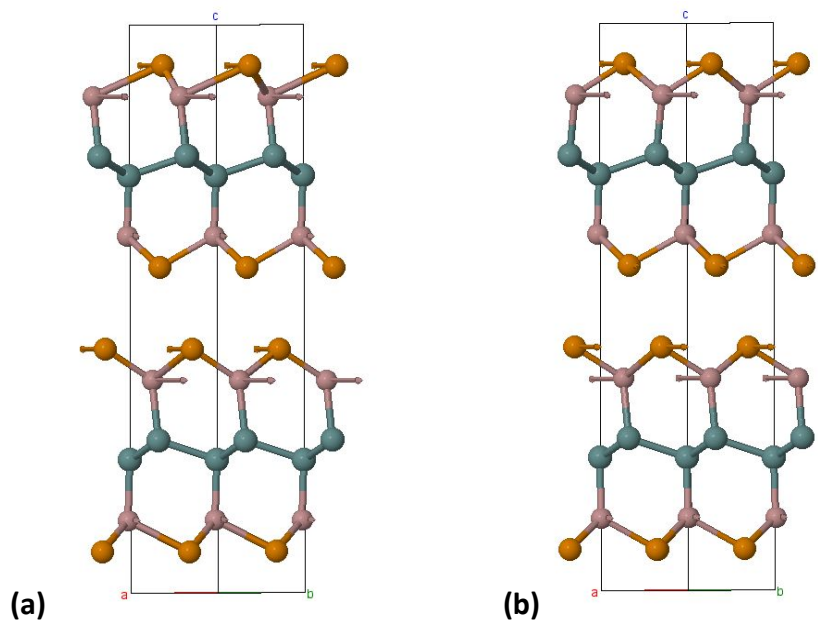

**Figure S17.** Atomic motion of $E_1^3$ (a) and $E_2^4$ (b) vibrational modes of β-GaGeTe predicted at 176.3 and 176.6 cm$^{-1}$. Ga, Ge, and Te atoms are depicted in pink, gray, and orange colors. These two modes are intra-layer modes of β-GaGeTe that are a mixture of a Ga-Ge bending mode and a Ga-Te stretching mode, where Ge atoms are at rest and Ga and Te atoms vibrate out-of pase in the a-b plane. In the $E_1^3$ mode, the same Ga and Te atoms of neighbour layers vibrate in-phase while in the $E_2^4$ mode, the same Ga and Te atoms of neighbour layers vibrate out-of-phase.

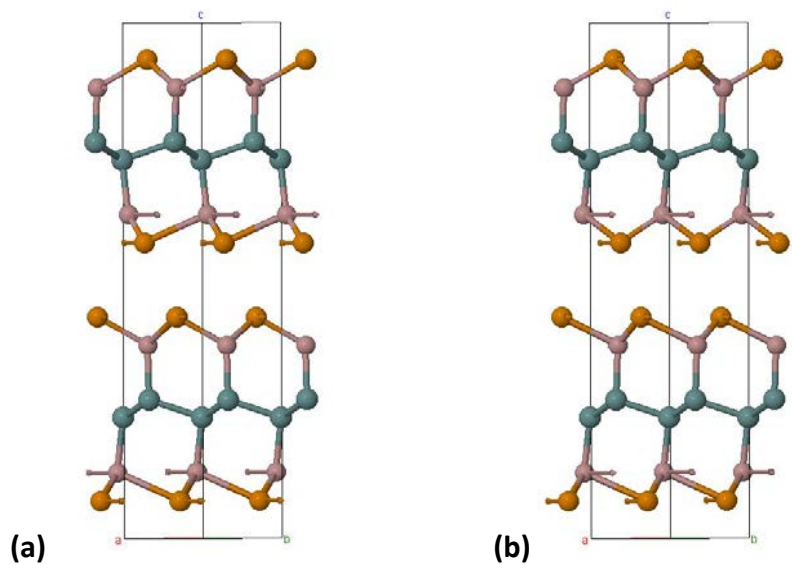

**Figure S18.** Atomic motion of $E_1^4$ (a) and $E_2^5$ (b) vibrational modes of β-GaGeTe predicted at 178.4 and 178.6 cm$^{-1}$. Ga, Ge, and Te atoms are depicted in pink, gray, and orange colors. These two modes are intra-layer modes of β-GaGeTe that are also a mixture of a Ga-Ge bending mode and a Ga-Te stretching mode, where Ge atoms are at rest and Ga and Te atoms vibrate out-of pase in the a-b plane. In the $E_1^4$ mode, the same Ga and Te atoms of neighbor layers vibrate in-phase while in the $E_2^4$ mode, the same Ga and Te atoms of neighbor layers vibrate out-of-phase. In the modes of Fig. S17 (Fig. S16), the amplitude of vibration is largest for the Ga and Te atoms of the bottom (top) of the layers.

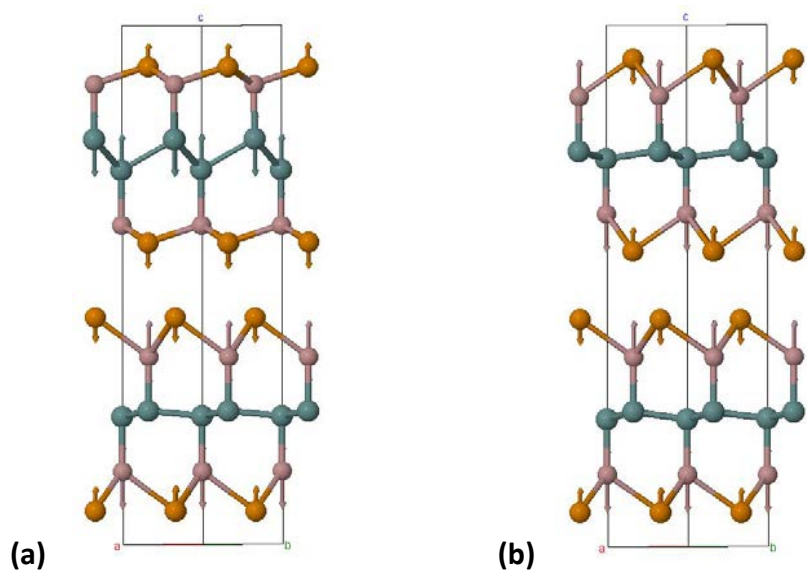

**Figure S19.** Atomic motion of $B_1^4$ (a) and $A_1^3$ (b) vibrational modes of β-GaGeTe predicted at 200.4 and 201.0 cm$^{-1}$. Ga, Ge, and Te atoms are depicted in pink, gray, and orange colors. These two modes are also longitudinal intra-layer modes of β-GaGeTe where Ge atoms vibrate in-phase along the c axis with neighbor Ga atoms in both bottom and top sublayers, which in turn vibrate out-of-phase with respect to Te atoms in their respective sublayers. Both modes are a mixture of a Ge-Ge bending mode and a Ga-Te bending mode. Additionally, the $A_1^3$ mode has a weak interlayer compressional mode due to the out-of-phase Te vibrations along the *c*-axis in the neighbor layers, that does not occur in the $B_1^4$ mode.

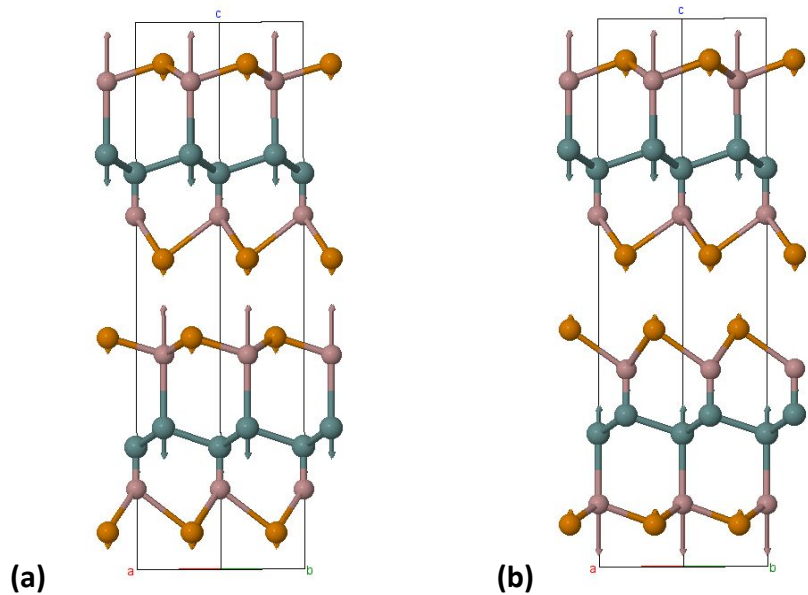

**Figure S20.** Atomic motion of $A_1^4$ (a) and $B_1^5$ (b) vibrational modes of β-GaGeTe predicted at 267.9 and 269.9 cm$^{-1}$. Ga, Ge, and Te atoms are depicted in pink, gray, and orange colors. These two modes are also longitudinal intra-layer modes of β-GaGeTe where Ge atoms vibrate out-of-phase along the c axis with respect to neighbor Ga atoms in both bottom and top sublayers, which in turn vibrate out-of-phase with respect to Te atoms in their respective sublayers. Both modes are a mixture of a Ge-Ge bending mode, a Ga-Te bending mode, and a Ga-Ge asymmetric stretching mode. In the $A_1^4$ mode, all atoms of one layer vibrate in-phase with those of the neighbor layer, while in the $B_1^5$ mode atoms of different layers vibrate out-of-phase.

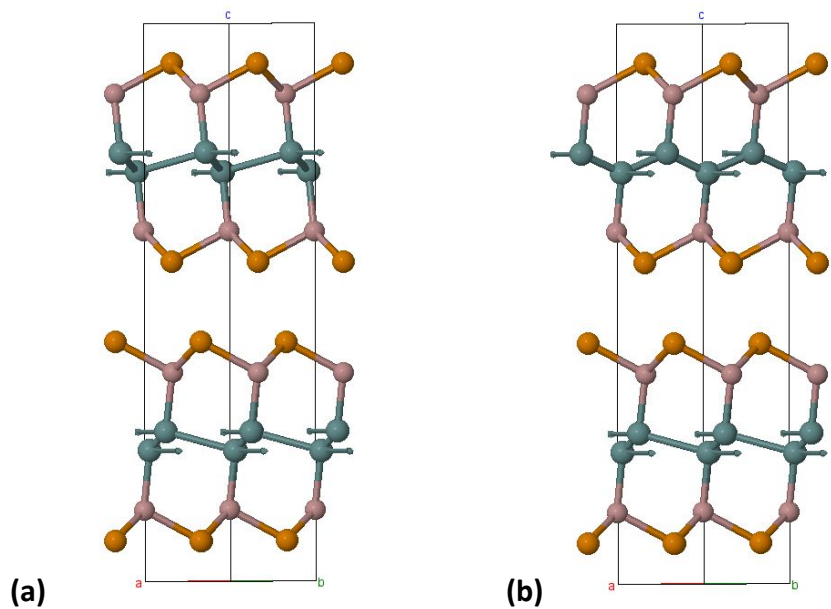

**Figure S21.** Atomic motion of $E_1^5$ (a) and $E_2^6$ (b) vibrational modes of β-GaGeTe predicted at 274.6 and 274.7 cm$^{-1}$. Ga, Ge, and Te atoms are depicted in pink, gray, and orange colors. These two modes show Te atoms at rest and Ga and Ge atoms in motion along the a-b plane, but the Ge atoms vibrate with much larger amplitude than Ga atoms. In both modes, neighbor Ge atoms vibrate out-of-phase. These modes are a mixture of a Ge-Ge asymmetric stretching mode and a Ga-Ge bending mode.

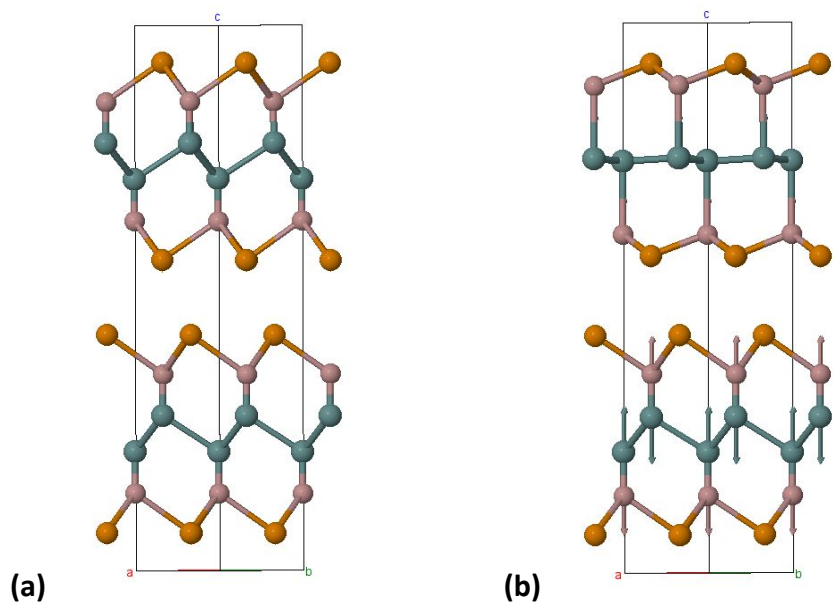

**Figure S22.** Atomic motion of $A_1^5$ (a) and $B_1^6$ (b) vibrational modes of $\beta$-GaGeTe predicted at 285.6 and 288.3 cm$^{-1}$. Ga, Ge, and Te atoms are depicted in pink, gray, and orange colors. These two modes are also longitudinal intra-layer modes of $\beta$-GaGeTe where Te atoms are at rest and neighbor Ge and Ga atoms vibrate out-of-phase along the c axis. These two modes are a mixture of a Ge-Ge bending mode, a Ga-Te bending mode and a Ga-Ge asymmetric stretching. In $A_1^5$ the vibrations in the two layers of the unit cell are in pase, while in $B_1^6$ the vibrations of the two layers are out-of-phase.

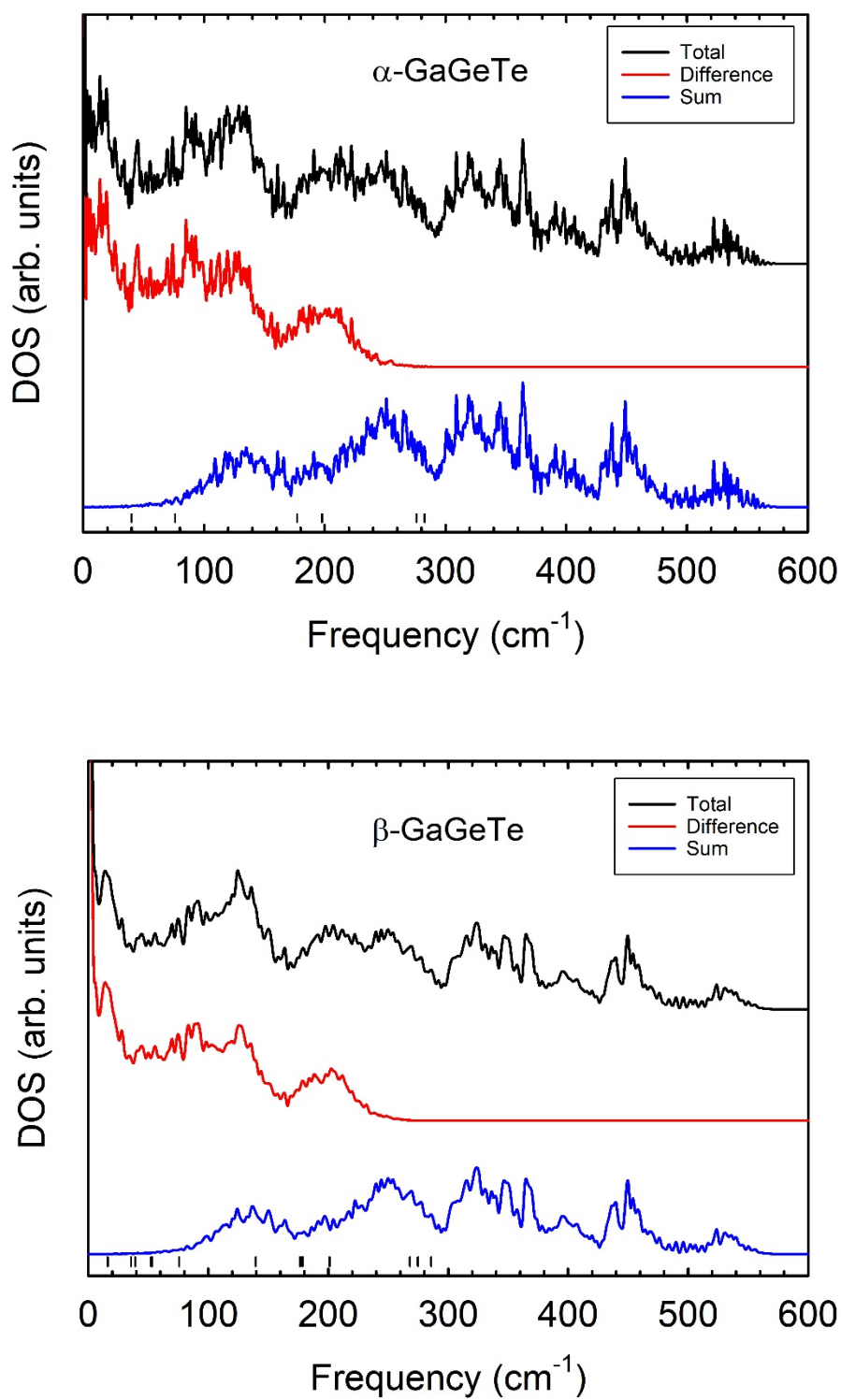

**Figure S23.** Theoretical total and partial two-phonon DOS of bulk α-GaGeTe (top) and bulk β-GaGeTe (bottom) at room conditions. Total, sum, and difference two-phonon DOS are plotted in black, blue, and red, respectively. The theoretical frequencies of the six first-order Raman-active modes of bulk α-GaGeTe and of the sixteen first-order Raman-active modes of bulk β-GaGeTe are represented as black bottom marks.

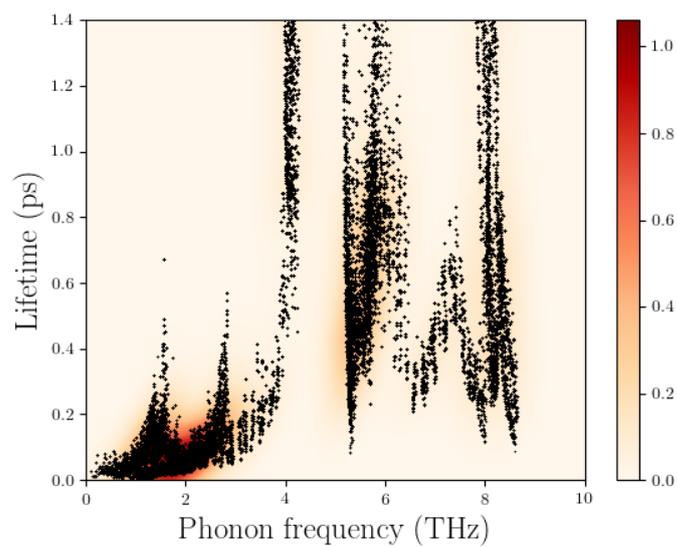

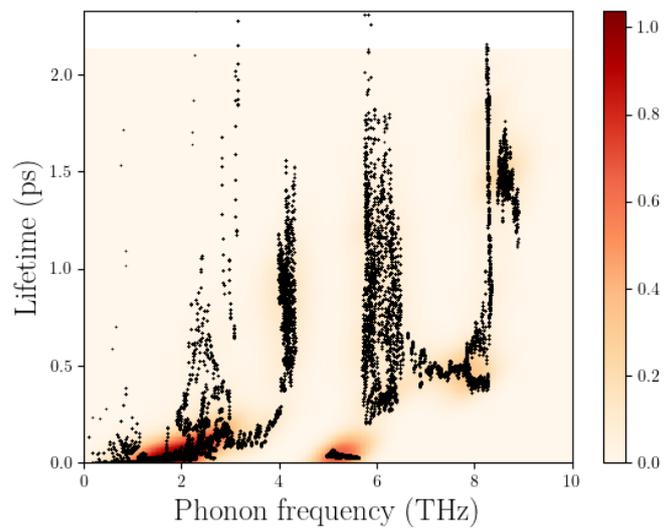

**Figure S24.** Calculated phonon lifetimes of α-GaGeTe (top) and β-GaGeTe (bottom) at 300 K. The color shades represent the phonon density, consequently darker shades refer to higher phonon densities.

**Table S1.** Theoretical PBEsol (PBE-D3) atomic coordinates of the *P-3m1* (s.g. No. 164) structure of GaGeTe-mono. The theoretical PBEsol (theoretical PBE-D3) lattice parameters of the hexagonal unit cell are: *a* = 4.0401 Å (4.0833 Å), *b* = 4.0401 Å (4.0833 Å), *c* = 12.4506 Å (12.3275 Å), and $V_0$ = 176.0000 Å$^3$ (178.0000 Å$^3$).

| Atom | Wyckoff site | x | y | z |
|---|---|---|---|---|
| Ga | 2d | 1/3 | 2/3 | 0.2729 (0.2689) |
| Ge | 2d | 1/3 | 2/3 | 0.4678 (0.4673) |
| Te | 2c | 0 | 0 | 0.1708 (0.1660) |

**Table S2.** Theoretical (PBEsol and PBE-D3) Raman- and infrared-active frequencies, ω, at ambient conditions for GaGeTe-mono. Theoretical (LDA) data from Ref. S3 for the α-GaGeTe monolayer (also with s.g. *P-3m1*) are also included for comparison.

| Mode | ω (th.)$^a$ (cm$^{-1}$) | ω (th.)$^b$ (cm$^{-1}$) | ω (th.)$^c$ (cm$^{-1}$) |
|---|---|---|---|
| $E_g^1$ | 35.6 | 34.4 | 39.0 |
| $E_u^1$ | 53.8 | 53.8 | 57.1 |
| $A_{1g}^1$ | 73.8 | 77.6 | 75.9 |
| $A_{2u}^1$ | 140.1 | 138.5 | 149.0 |
| $E_g^2$ | 177.2 | 172.9 | 186.9 |
| $E_u^2$ | 177.3 | 173.2 | 186.9 |
| $A_{1g}^2$ | 205.5 | 203.4 | 220.0 |
| $A_{2u}^2$ | 269.4 | 266.4 | 284.7 |
| $E_g^3$ | 277.1 | 270.5 | 284.7 |
| $A_{1g}^3$ | 291.1 | 287.7 | 305.2 |

$^a$ PBEsol, $^b$ PBE-D3, $^c$ Estimated from Ref. S3.

**Table S3.** Theoretical PBEsol (PBE-D3) atomic coordinates of the *P-3m1* (s.g. No. 164) structure of γ-GaGeTe. The theoretical PBEsol (theoretical PBE-D3) lattice parameters of the hexagonal unit cell are: *a* = 4.0220 Å (4.0730 Å), *b* = 4.0220 Å (4.0730 Å), *c* = 11.5638 Å (11.6937 Å), and $V_0$ = 162.0000 Å$^3$ (168.0000 Å$^3$).

| Atom | Wyckoff site | x | y | Z |
|---|---|---|---|---|
| Ga | 2d | 1/3 | 2/3 | 0.7545 (0.7553) |
| Ge | 2d | 1/3 | 2/3 | -0.0351(-0.0348) |
| Te | 2d | 1/3 | 2/3 | 0.3567 (0.3540) |

**Table S4.** Theoretical (PBEsol and PBE-D3) Raman- and infrared-active frequencies, ω, at ambient conditions for γ-GaGeTe. Theoretical (LDA) data from Ref. S3 for the α-GaGeTe monolayer (also with s.g. *P-3m1*) are also included for comparison.

| Mode | ω (th.)$^a$ (cm$^{-1}$) | ω (th.)$^b$ (cm$^{-1}$) | ω (th.)$^c$ (cm$^{-1}$) |
|---|---|---|---|
| $E_g^1$ | 37.7 | 38.1 | 39.0 |
| $E_u^1$ | 52.4 | 53.0 | 57.1 |
| $A_{1g}^1$ | 79.0 | 79.7 | 75.9 |
| $A_{2u}^1$ | 142.3 | 140.6 | 149.0 |
| $E_u^2$ | 177.3 | 173.5 | 186.9 |
| $E_g^2$ | 178.5 | 174.7 | 186.9 |
| $A_{1g}^2$ | 205.3 | 203.4 | 220.0 |
| $A_{2u}^2$ | 268.1 | 264.3 | 284.7 |
| $E_g^3$ | 278.1 | 270.5 | 284.7 |
| $A_{1g}^3$ | 289.6 | 285.3 | 305.2 |

[a] PBEsol, [b] PBE-D3, [c] Estimated from Ref. S3.